\documentclass[superscriptaddress,aps,prd,showpacs,preprint]{revtex4-1}

\usepackage{eurosym}
\usepackage{graphicx}
\usepackage{amsmath}
\usepackage{mathrsfs}
\usepackage{bm}
\usepackage{color}
\usepackage{amsmath}
\usepackage{amssymb}

\usepackage{multirow}
\usepackage{array}
\usepackage{subfig}
\newcommand{\bea}{\begin{eqnarray}}
\newcommand{\ena}{\end{eqnarray}}
\newcommand{\be}{\begin{equation}}
\newcommand{\ee}{\end{equation}}

\allowdisplaybreaks

\begin{document}

\title{Quasinormal Modes of Near-Extremal Black Holes and Black Strings in Massive Gravity Background}
\author{Taum Wuthicharn}
\email{Taum.W5@gmail.com}
\affiliation{High Energy Physics Theory Group, Department of Physics,
	Faculty of Science, Chulalongkorn University, Phyathai Rd., Bangkok 10330, Thailand}
\author{Supakchai Ponglertsakul}
\email{Supakchai.p@gmail.com}
\affiliation{High Energy Physics Theory Group, Department of Physics,
	Faculty of Science, Chulalongkorn University, Phyathai Rd., Bangkok 10330, Thailand}
\author{Piyabut Burikham}
\email{piyabut@gmail.com}
\affiliation{High Energy Physics Theory Group, Department of Physics,
	Faculty of Science, Chulalongkorn University, Phyathai Rd., Bangkok 10330, Thailand}

\date{\today }

\begin{abstract}
	
	Two numerical methods are used to calculate quasinormal modes~(QNMs) of near-extremal black holes/strings in the generalized spherically/cylindrically symmetric background, the Asymptotic Iteration Method~(AIM) and the Spectral Method.  The numerical results confirm the accuracy of the approximate analytic formula using the P\"{o}schl-Teller potential. Our analytic formula is used to investigate the Strong Cosmic Censorship conjecture of extremal and near-extremal black holes in Ref.~\cite{Burikham:2020dfi}.
	
	\vspace{5mm}
	
	{Keywords: Quasinormal Modes, Extremal Black Hole, Massive gravity, Black String, AIM, Spectral Method}
	
\end{abstract}

\maketitle

\section{Introduction}

Isolated black holes in equilibrium are very simple objects. Only three parameters are needed to uniquely describe them i.e. mass, angular momentum and charge. However, black holes are scarcely isolated. They will eagerly interact with their surroundings e.g. fields, particles, accretion disk and strong magnetic field. At a fundamental level, interaction with black hole can be treated as a perturbation on the black hole. When perturbed, black hole responds by releasing gravitational waves which carry encoded information in their oscillatory modes unique to the sources. 

The first gravitational event has been detected by the Laser Interferometer Gravitational-Wave Observatory~(LIGO) in $2015$ \cite{Abbott:2016blz} . The merging of two stellar mass black holes emits sufficiently strong gravitational wave signal for LIGO. The signal after the merging is exponentially decayed. This damping oscillation phase is described by quasinormal modes (QNMs). The associated frequencies of QNMs are quasinormal frequencies which are discrete and complex. An oscillation of quasinormal frequency is characterized by its real part while  damping oscillation is related to its imaginary part \cite{Kokkotas:1999bd}. Black hole mass and spin can be deduced from the characteristic decaying time of gravitational waves. In addition, several studies have been proposed to explore a characteristic of gravitational waves from wormhole and other compact objects \cite{Kim:2008zzj, Konoplya:2016hmd, Cardoso:2016rao}. The VIRGO interferometer later joined LIGO in 2017 \cite{Abbott:2017oio} has marked the beginning of the gravitational wave astronomy era.

The evolution of wavepacket in the fixed Schwarzschild background was studied in 1970 by Vishveshwara \cite{vivesh}. It turns out that the signal is dominated by the damped oscillation. Since then, a vast number of studies has been devoted to the QNMs of black holes in various asymptotic backgrounds i.e., asymptotically flat,  de-Sitter (dS) and Anti de-Sitter (AdS) (well-written reviews on this subject can be found in \cite{Kokkotas:1999bd, Berti:2009kk, Konoplya:2011qq}). A special kind of black hole configuration that has intriguing properties is an extremal black hole. In general, extremal black hole has zero surface gravity and thus zero temperature. Hence, when extremal black hole reaches its extremal limit the evaporation process will stop. According to Bekenstein-Hawking entropy \cite{Bekenstein:1972tm, Hawking:1974sw}, black hole's horizon area is associated to its entropy. However, an extremal Reissner-Nordstr\"om black hole with non-vanishing horizon area still possess zero entropy  \cite{Hawking:1994ii}. Charged and rotating extremal BTZ black holes are shown to be stable against small perturbation \cite{Crisostomo:2004hj, Richartz:2015saa}. Moreover, various investigations on the QNMs of extremal/near-extremal black holes have been performed. For instance, the highly damped QNMs of extremal Reissner-Nordstr\"om dS black holes are investigated in Ref.~\cite{Daghigh:2007xj}. Gravitational perturbation of higher dimensional Reissner-Nordstr\"om dS black holes is explored in the near extremal limit \cite{Konoplya:2008au} where black holes in $D\geq 7$ spacetime dimensions are unstable. An exact relation for QNMs of scalar perturbations on near-extremal black holes in Weyl gravity is explored in \cite{Momennia:2019cfd}. QNMs of near-extremal Schwarzschild dS black holes are obtained analytically via the P\"oschl-Teller technique \cite{Cardoso:2003sw, MaassenvandenBrink:2003yq} and numerically by Yoshida and Futamase in Ref.~\cite{Yoshida:2003zz}. Scalar perturbations on $d$ dimensional Myper-Perry de Sitter black hole with single rotation and their quasinormal spectrum are also investigated in the near-extremal limit \cite{Ponglertsakul:2020ufm}. In theory beyond general relativity, QNMs of near-extremal black string in de Rham-Gabadadze-Tolley (dRGT) massive gravity theory are studied both analytically and numerically \cite{Ponglertsakul:2018smo}.

The QNMs problem usually reduces to solving second-order differential equation with appropriate boundary conditions. Due to the complexity of the spacetime background, the QNMs are normally obtained via numerical methods. However, there are some~(semi-) analytical approaches to calculate the QNMs. For instances, matching asymptotic solutions are used in several works related to black hole perturbations including QNMs \cite{Page:1976df,Unruh:1976fm,Cardoso:2004nk,Li:2014gfg}. The Wentzel-Kramers-Brillouin (WKB) approximation is also widely implemented in various QNMs studies \cite{Ponglertsakul:2020ufm,Ponglertsakul:2018smo,Iyer:1986np,Konoplya:2003ii,Burikham:2017gdm}. In the near-extremal limit, the effective potential of radial wave equation can be expressed in terms of the P\"oschl-Teller potential \cite{Poschl1933} where the exact solution is known. The quasinormal spectrum can thus be written out explicitly \cite{Cardoso:2003sw, Ponglertsakul:2020ufm, Ponglertsakul:2018smo, Ferrari:1984zz}. 

In the previous work \cite{Ponglertsakul:2018smo}, the QNMs of scalar perturbation on neutral dRGT black string in the near-extremal limit with positive cosmological constant where the event horizon approaches the cosmic horizon~(the Nariai limit) have been explored. The P\"oschl-Teller technique is applied and an analytic formula for the quasinormal spectrum is obtained. In this work, we consider near-extremal limit in all possible scenarios (not only in the Nairai limit) of charged dRGT black holes and neutral black strings in asymptotically dS and AdS. As described in details in Ref.~\cite{Burikham:2020dfi}, we find that the approximation can be extended to more generic cases of near-extremal black hole spacetime where the Cauchy horizon approaching the event horizon but remotely separated from the cosmic horizon, i.e., large universe scenario. To verify our analytic results, we compare them with the results from AIM and the spectral method in this work. It turns out that equation of motion for spin-0 perturbation is similar to spin-1 case. Therefore the results from the former is also applicable to the latter. Numerical results in this work are found to be in exact agreement with the approximate analytic formula, Eq.~(\ref{QNMF}).  Remark that, we also use  the method described in this paper to explore Strong and Weak Cosmic Censorship conjecture of various black holes in \cite{Burikham:2020dfi}.

This paper is organized as follows. In Section \ref{setup}, we explore a number of scenarios of extremal black holes whose background parameters are slightly altered to create near-extremal black holes. The analytic formula~(derived in Ref.~\cite{Burikham:2020dfi}) of the near-extremal QNMs of charged black holes in the massive gravity background is presented in Section \ref{Math}. The numerical analyses using AIM and the spectral method are performed and the comparison to the analytic values are presented in Section \ref{dSN} and \ref{AdSN}. In Section \ref{BS} we calculate the QNMs of near-extremal black string in dS and AdS spacetime using the same techniques. Spin-1 perturbation is demonstrated to obey the same analytic formula in Section~\ref{Sone}. Section \ref{conclude} concludes our work.

\section{Extremal Black Holes in Massive Gravity Background}\label{setup}

In this Section, we will explore various types of extremal black hole. The extremal black hole parameters are then slightly shifted to obtain the near-extremal black hole. The parameter space has been studied to certain extent in Ref.~\cite{Burikham:2020dfi}, here we will explore it in more details. 

The metric of the massive gravity can be described as follows,
	\begin{equation}
		ds^{2}=-f(r)dt^{2}+\frac{dr^{2}}{f(r)}+r^{2}d\Omega^{2},\nonumber
	\end{equation}
where
	\begin{equation}\label{gmet}
		f(r)=1+\epsilon_{0}-\frac{2M}{r}+\frac{Q^{2}}{r^{2}}+\gamma r-\frac{\Lambda}{3}r^{2}.
	\end{equation}
The above form is valid for a number of different types of massive gravity models~\cite{Hinterbichler:2011tt, Burikham:2016roo}. The metric $f(r)$ has four possible roots. If these roots are real and positive, then they will represent horizons. It is known that there are three types of horizon. A Cauchy Horizon ($r_{C}$) is an inner horizon of the black hole; it covers the singularity. An Event Horizon which will be denoted by $r_{H}$. A Cosmic Horizon ($r_{\Lambda}$) is a boundary where the expansion of the Universe will prevent any signals originated beyond it to reach the physical Universe.

The $f(r)$ contains five different parameters $\epsilon_{0}$, $M$, $Q$, $\gamma$ and $\Lambda$. For the sake of simplicity, the parameter $\epsilon_{0}$ is set to zero. Moreover, it is possible to rescale other parameter around the parameter $M$. A method which preserves the generality while $M$ is set to unity (See Appendix \ref{ReS} for more detail). To further simplify, we separate our problem into two cases. The asymptotically de-Sitter ($\Lambda>0$) and Anti de-Sitter ($\Lambda<0$) cases. The asymptotically flat case is not considered.

\subsection{Extremal Black Hole in Asymptotic de-Sitter Case}

By studying the behavior of $f(r)$ for $\Lambda>0$, it is observed that $f(r)$ always has two different real roots, a negative  $r_{-}$ and a positive $r_{\Lambda}$~(cosmic horizon). The other two roots can either be complex conjugate of one another or both real depending on $\gamma$, $Q$ and $\Lambda$. These latter two roots are denoted $r_{C}$ and $r_{H}$ if they are real.

\begin{figure}[h]
		\centering
		\subfloat[]{\includegraphics[width=0.3\textwidth]{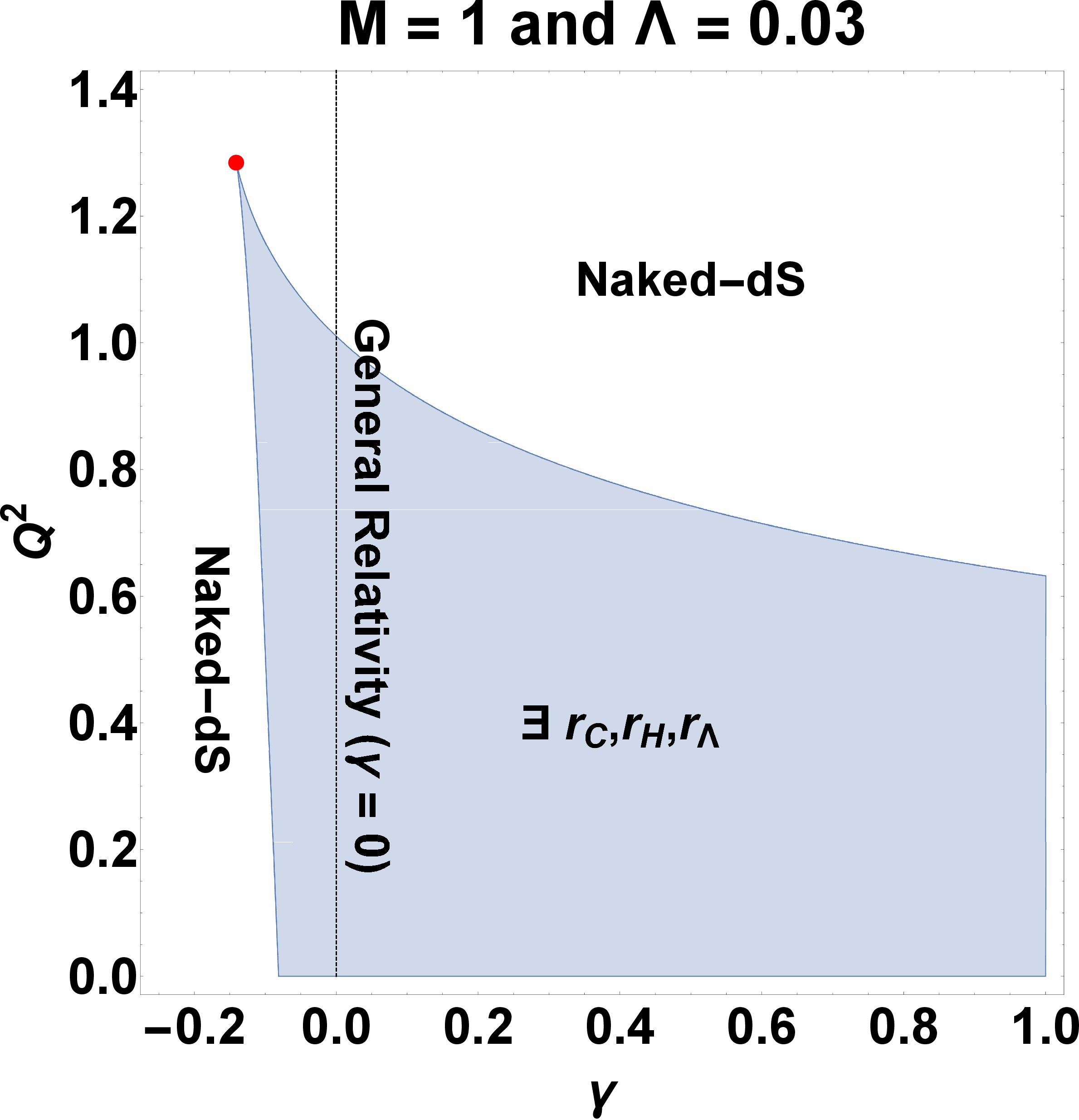}}
		\hspace{1em}
		\subfloat[]{\includegraphics[width=0.3\textwidth]{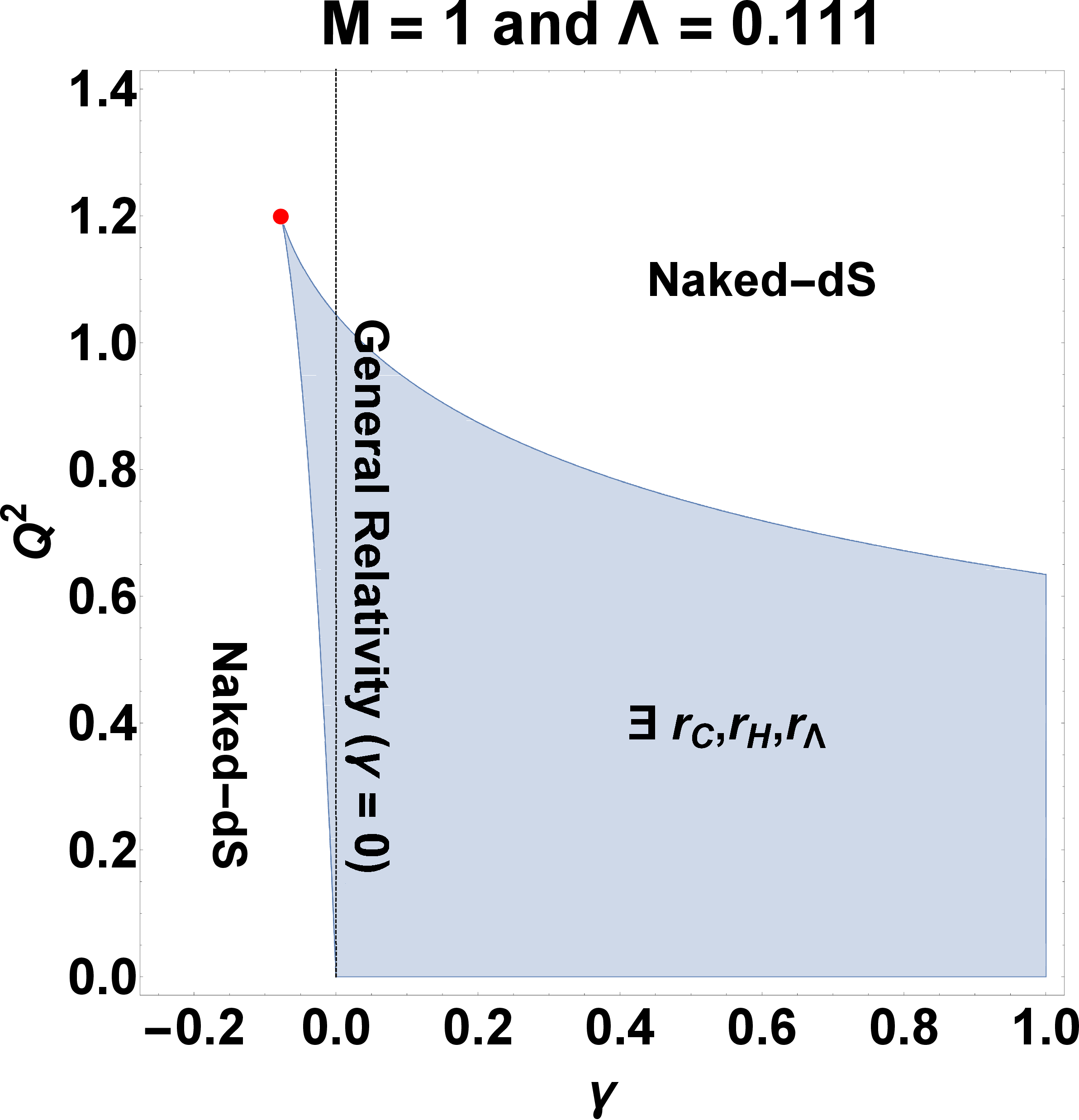}}
		\hspace{1em}
		\subfloat[]{\includegraphics[width=0.3\textwidth]{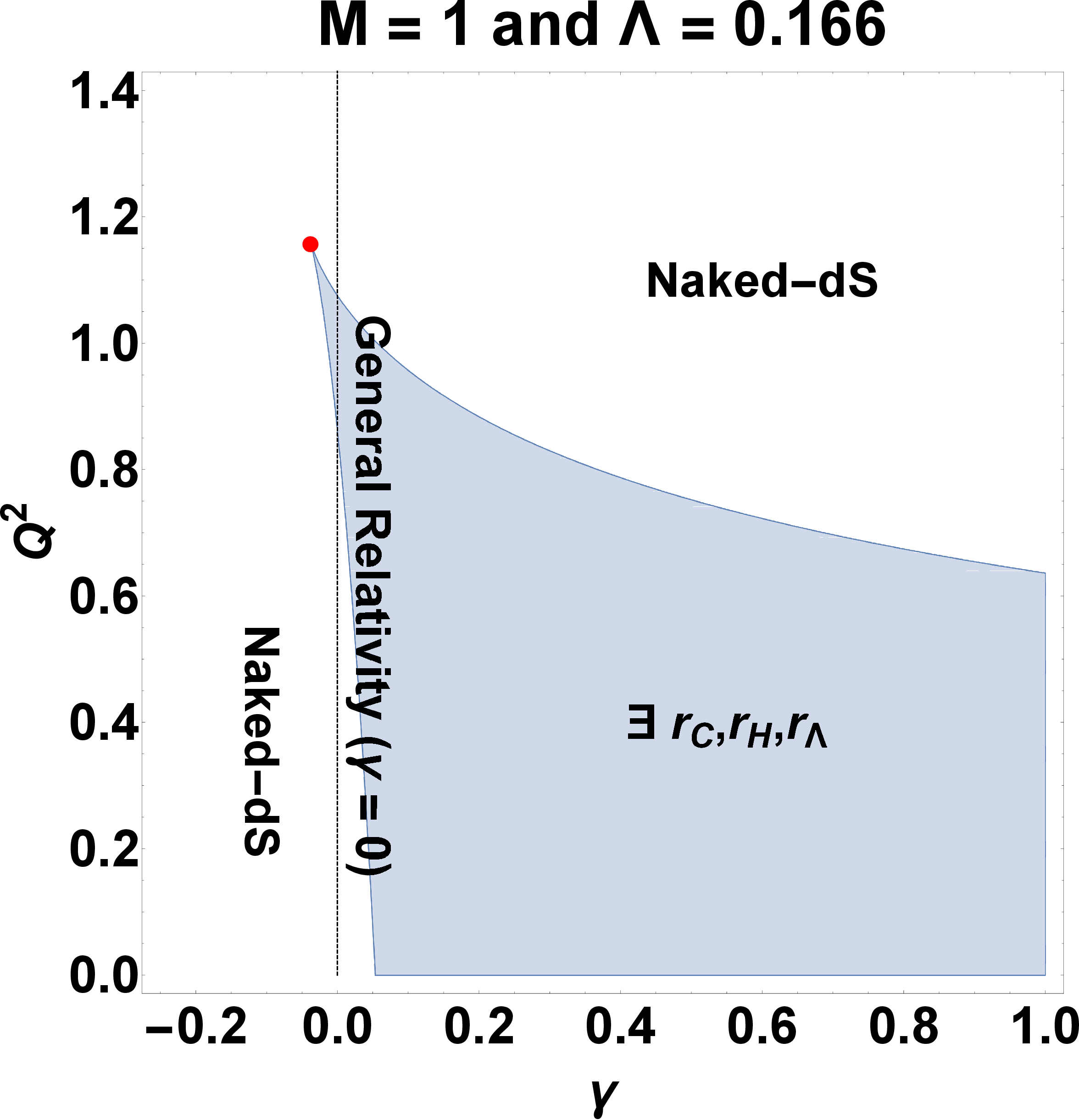}} \\
		\subfloat[]{\includegraphics[width=0.3\textwidth]{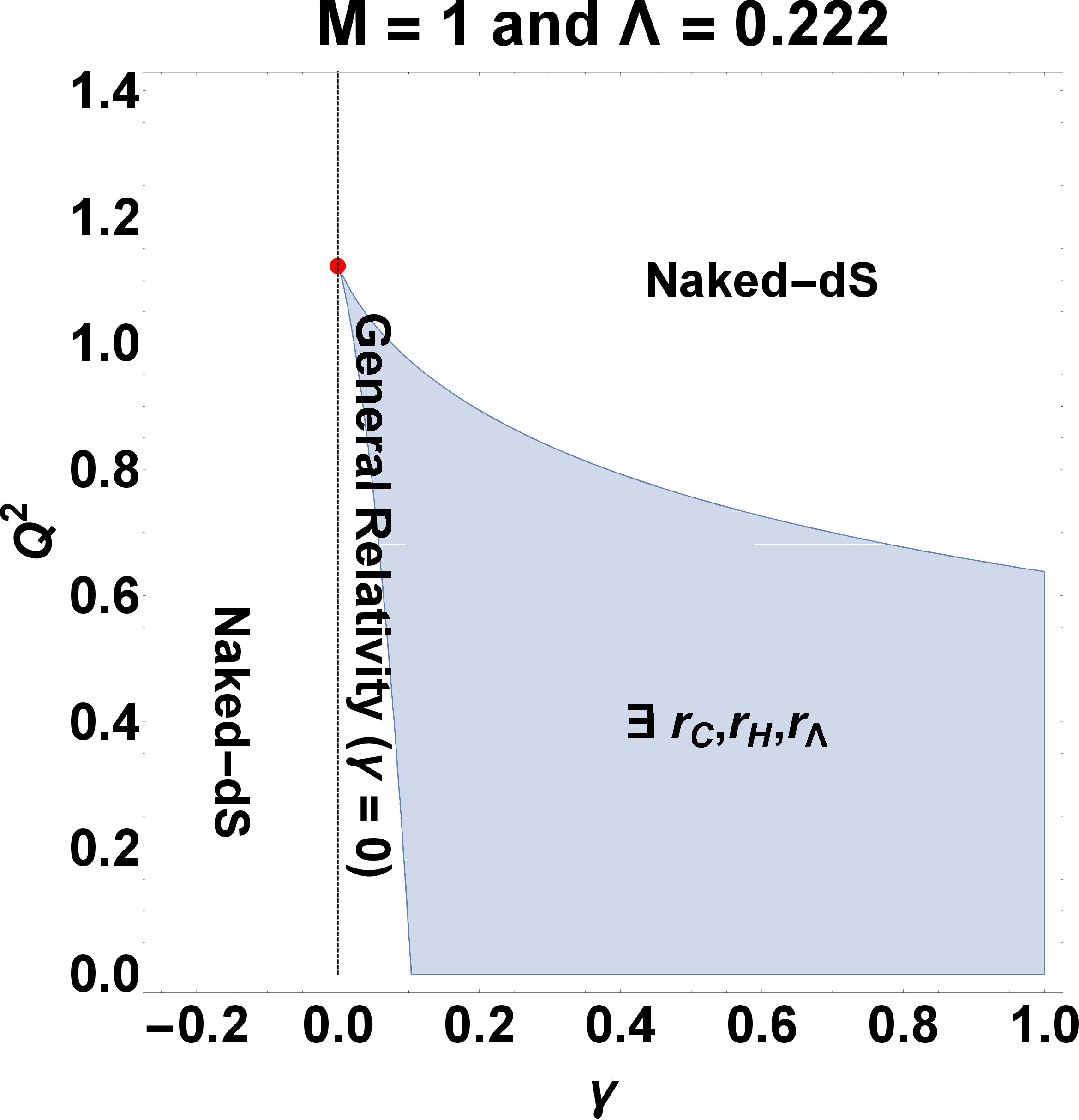}}
		\hspace{2em}
		\subfloat[]{\includegraphics[width=0.3\textwidth]{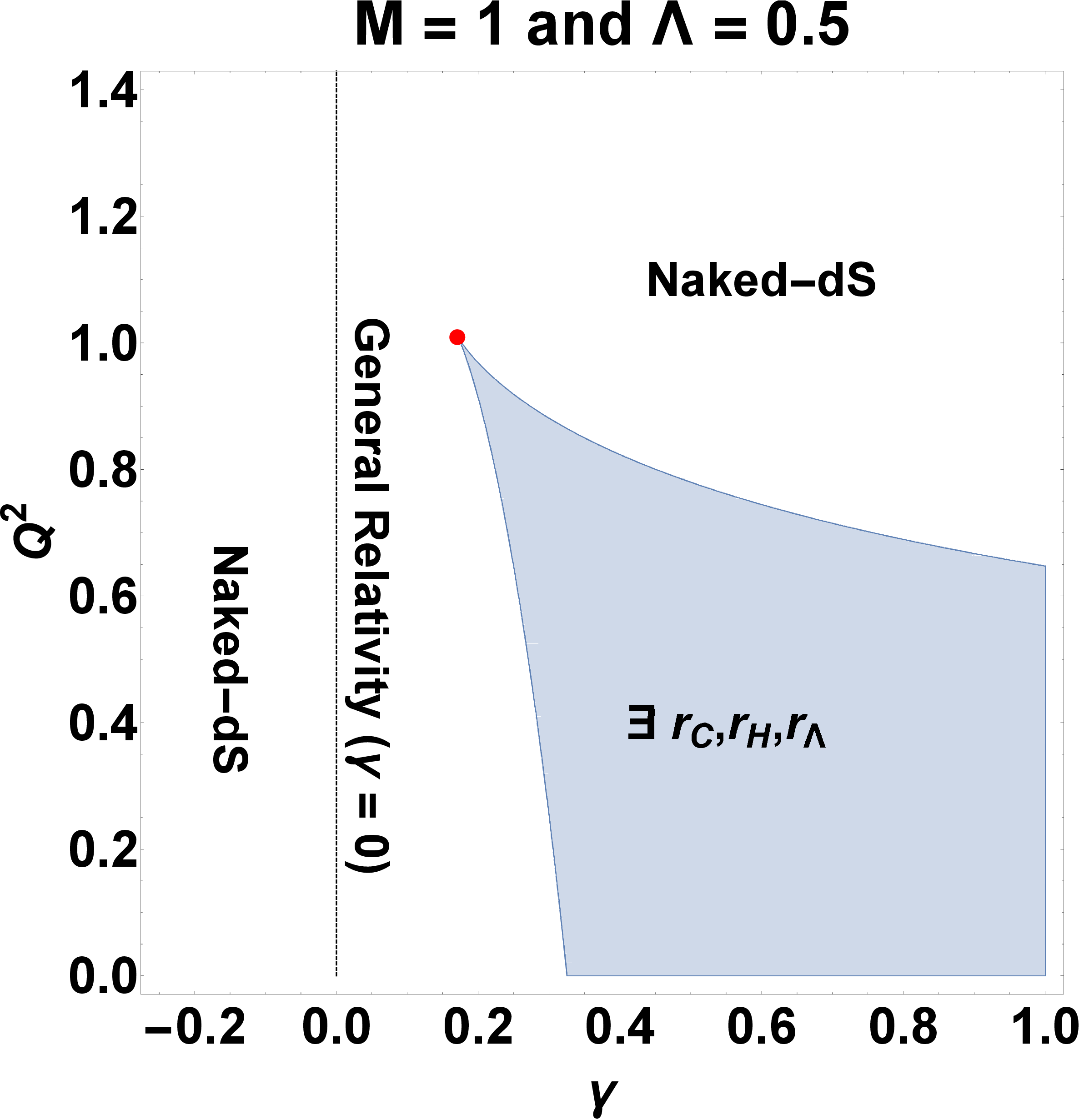}}
		\caption{Parameter space in $\gamma$-$Q^{2}$ with positive $\Lambda$.}
		\label{Sample-dS}
	\end{figure}

By setting $\Lambda$ to a number of different value, $f(r)$ is investigated in $\gamma$-$Q^{2}$ parameter space. The results are shown in FIG. \ref{Sample-dS}.
The `$\exists~r_{C},r_{H},r_{\Lambda}$' region represents the region when the black hole possesses a Cauchy Horizon and an Event Horizon. The `Naked-dS' region represents the region where the physical Universe possesses a naked singularity. The line between each region is then an extremal locus. The curved vertical line corresponds to an extremal type $r_{H}\sim r_{\Lambda}$. This locus will be called `Small Universe' locus. The curved horizontal line corresponds to an extremal type $r_{C}\sim r_{H}$. This locus will be called `Large Universe' locus. Both lines converge at a red dot. This dot represents a double extremal black hole ($r_{C}\sim r_{H}\sim r_{\Lambda}$).

As shown in FIG. \ref{Sample-dS}, the region `$\exists~r_{C},r_{H},r_{\Lambda}$' moves to the right as $\Lambda$ increases. It is interesting to consider the following statements in General Relativity Limit ($\gamma=0$). For a case of $\Lambda<1/9$, a black hole can only be one type of extremal ($r_{C}\sim r_{H}$). At $\Lambda=1/9$, the `Small Universe' locus touches the origin, $(\gamma,Q^{2})=(0,0)$. This coincides with the Nariai limit, where the Schwarzschild de-Sitter black hole becomes extremal. For the case where $1/9<\Lambda<2/9$, a black hole can be any of the two types of extremal; depending on the value of $Q$. At $\Lambda=2/9$ and further, it will be impossible to cover the singularity with the Cauchy Horizon.

\subsection{Extremal Black Hole in Asymptotic Anti de-Sitter Case}

In this case with $\Lambda <0$, $f(r)$ may possess up to four real roots, $r_{C}, r_{H}, r_{\Lambda-}, r_{\Lambda+}$, and approaches $+\infty$ for $r\to \pm \infty$. $r_{\Lambda-}$ acts as a Cosmic Horizon for a physical Universe between $r_{H}$ and $r_{\Lambda-}$. This ``Island'' Universe behaves locally similar to a de-Sitter Universe. In case that the black hole has no $r_{C}$ and $r_{H}$, $r_{\Lambda-}$ will act as the Cauchy Horizon for an unbound physical Universe beyond $r_{\Lambda+}$. $r_{\Lambda+}$ will always act as an event horizon for the unbound Universe.

\begin{figure}[h]
		\centering
		\subfloat[]{\includegraphics[width=0.3\textwidth]{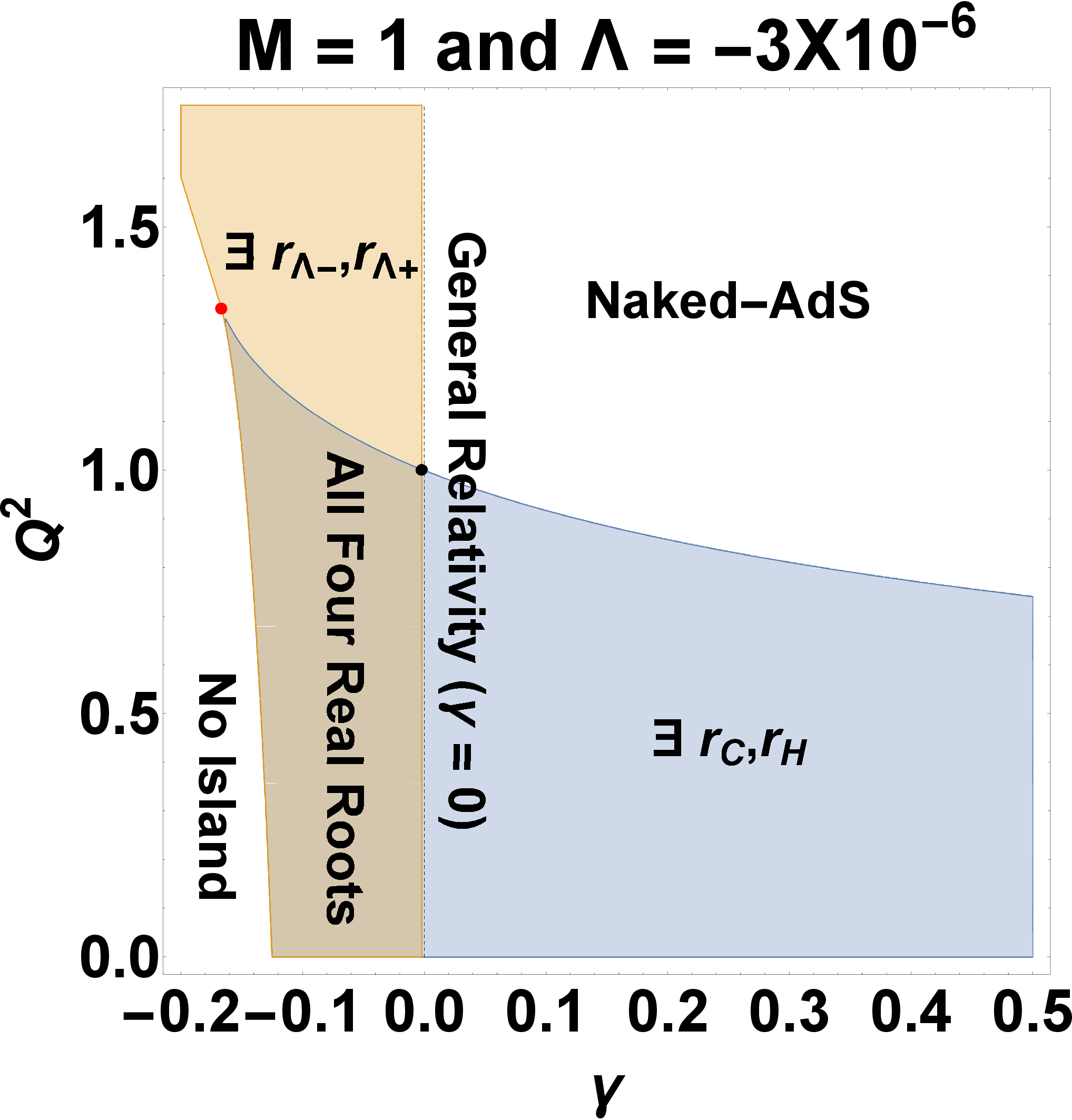}}
		\hspace{1em}
		\subfloat[]{\includegraphics[width=0.3\textwidth]{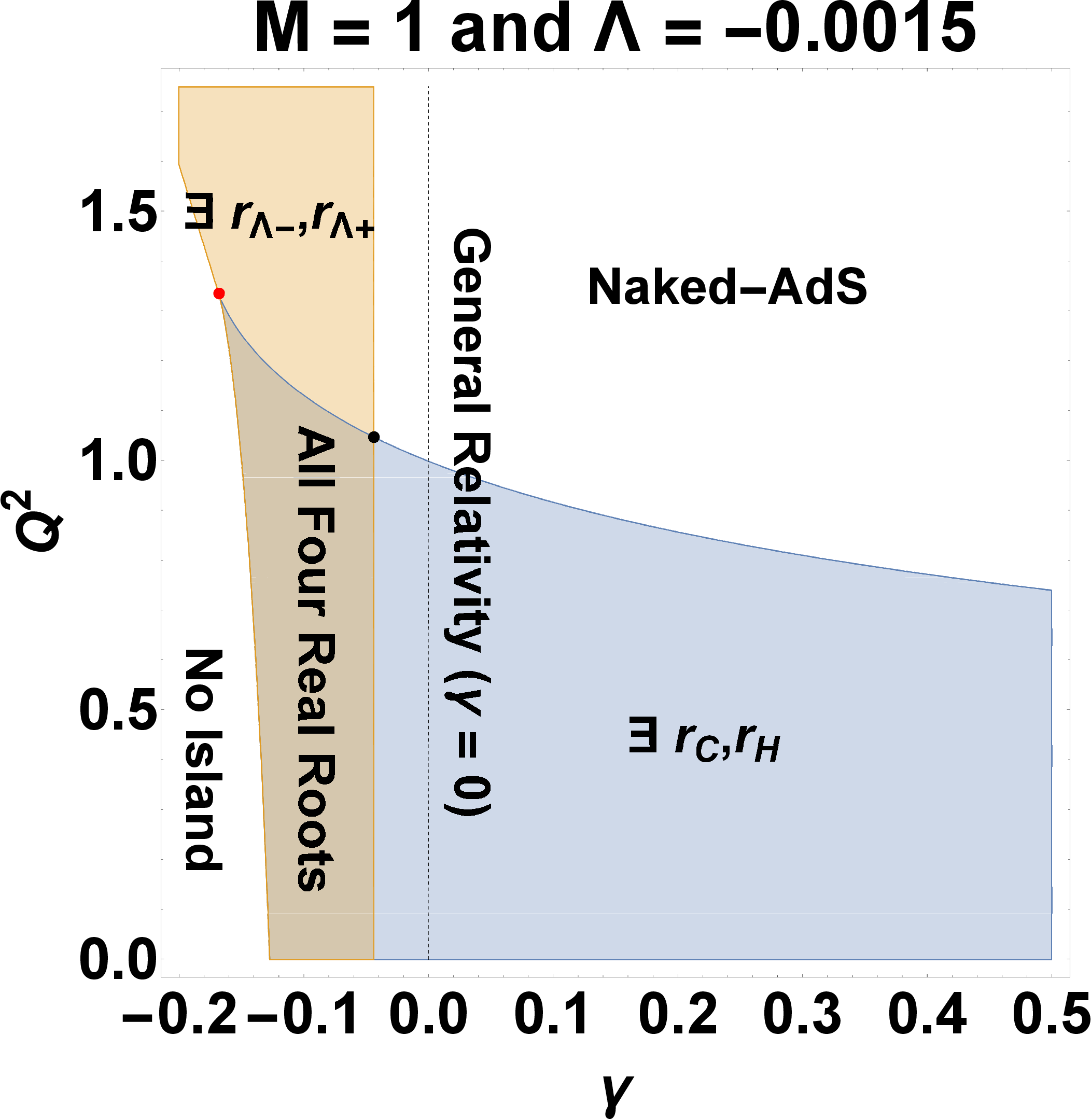}}
		\hspace{1em}
		\subfloat[]{\includegraphics[width=0.3\textwidth]{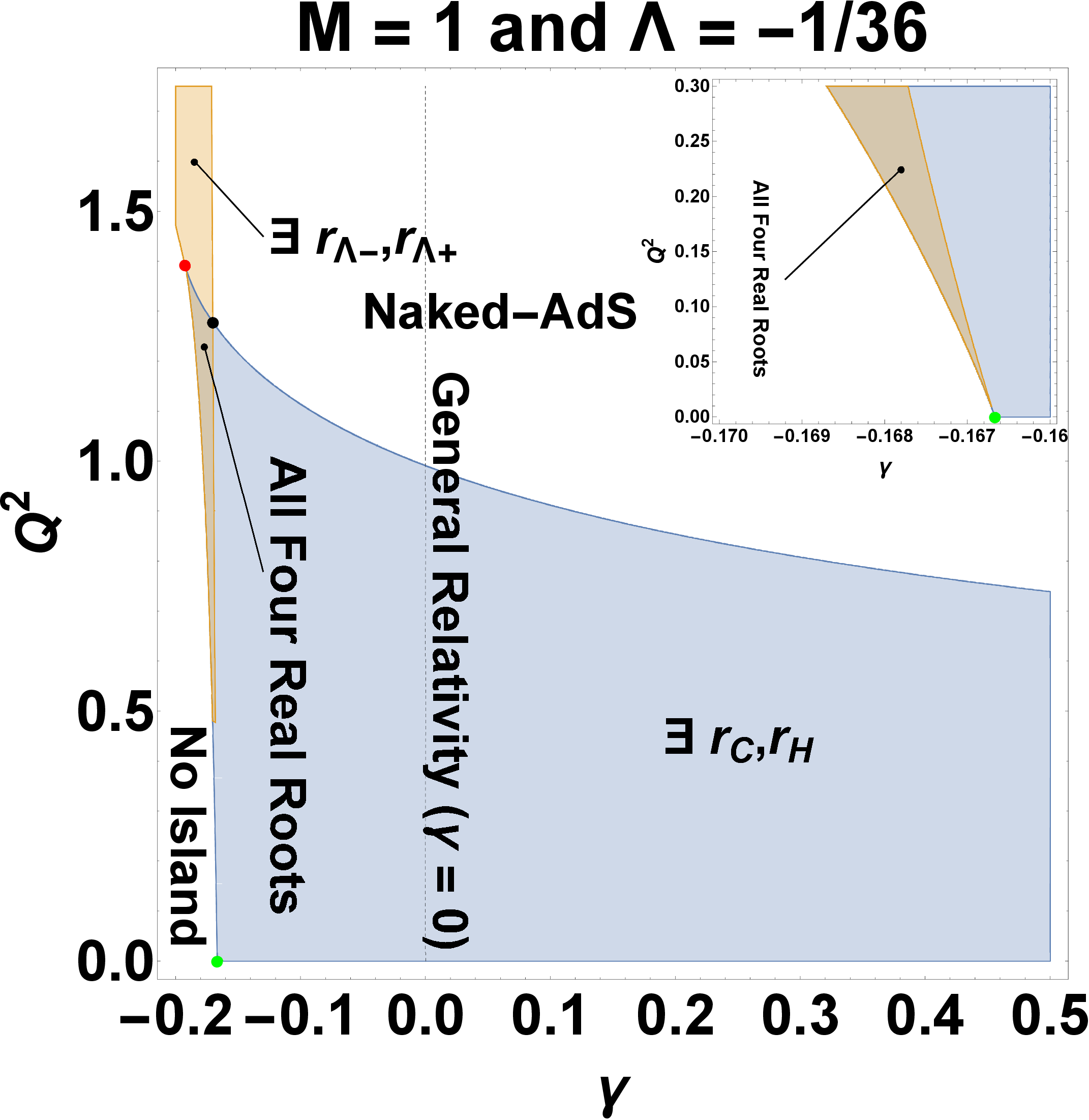}} \\
		\subfloat[]{\includegraphics[width=0.3\textwidth]{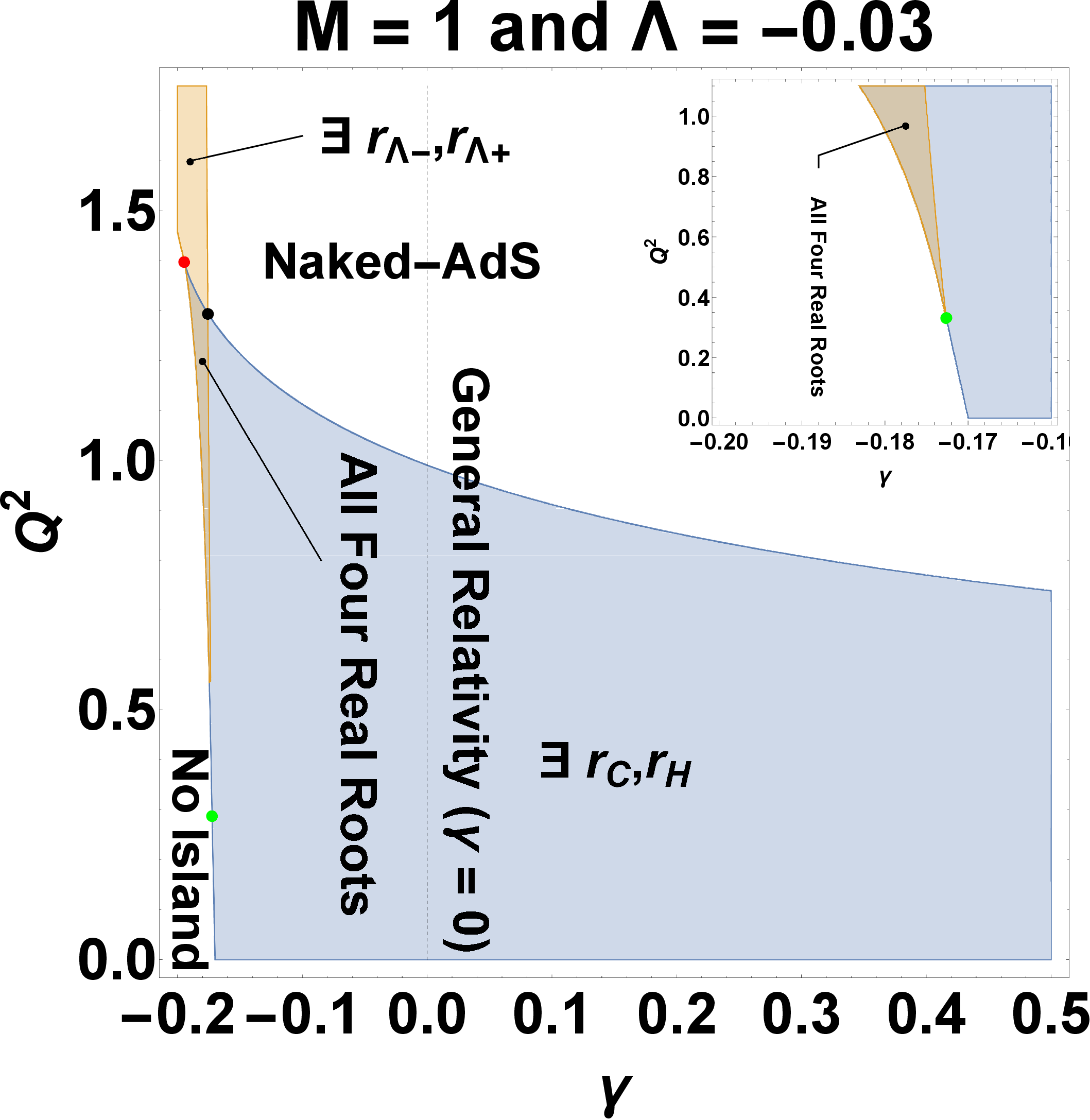}}
		\hspace{1em}
		\subfloat[]{\includegraphics[width=0.3\textwidth]{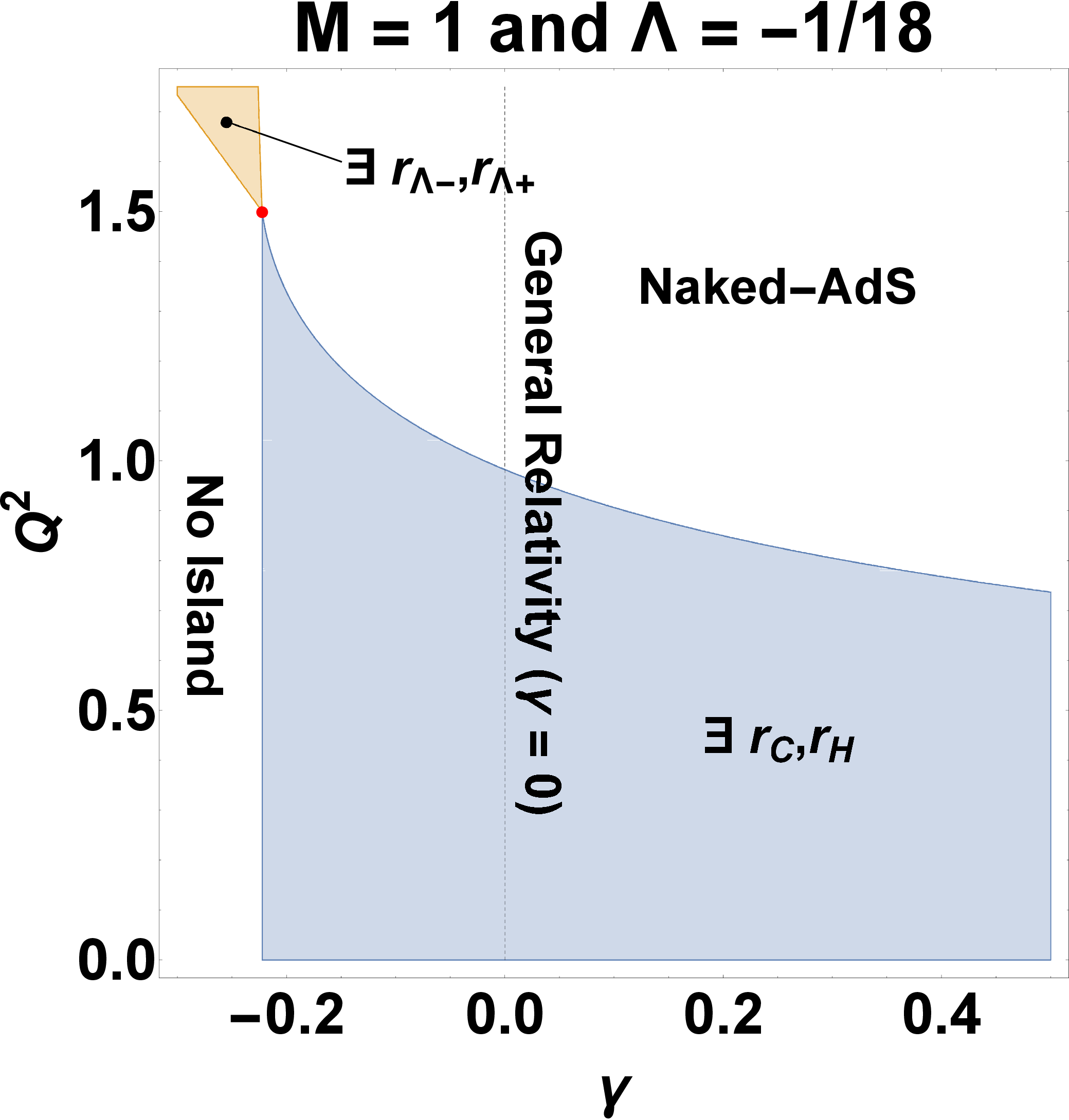}}
		\hspace{1em}
		\subfloat[]{\includegraphics[width=0.3\textwidth]{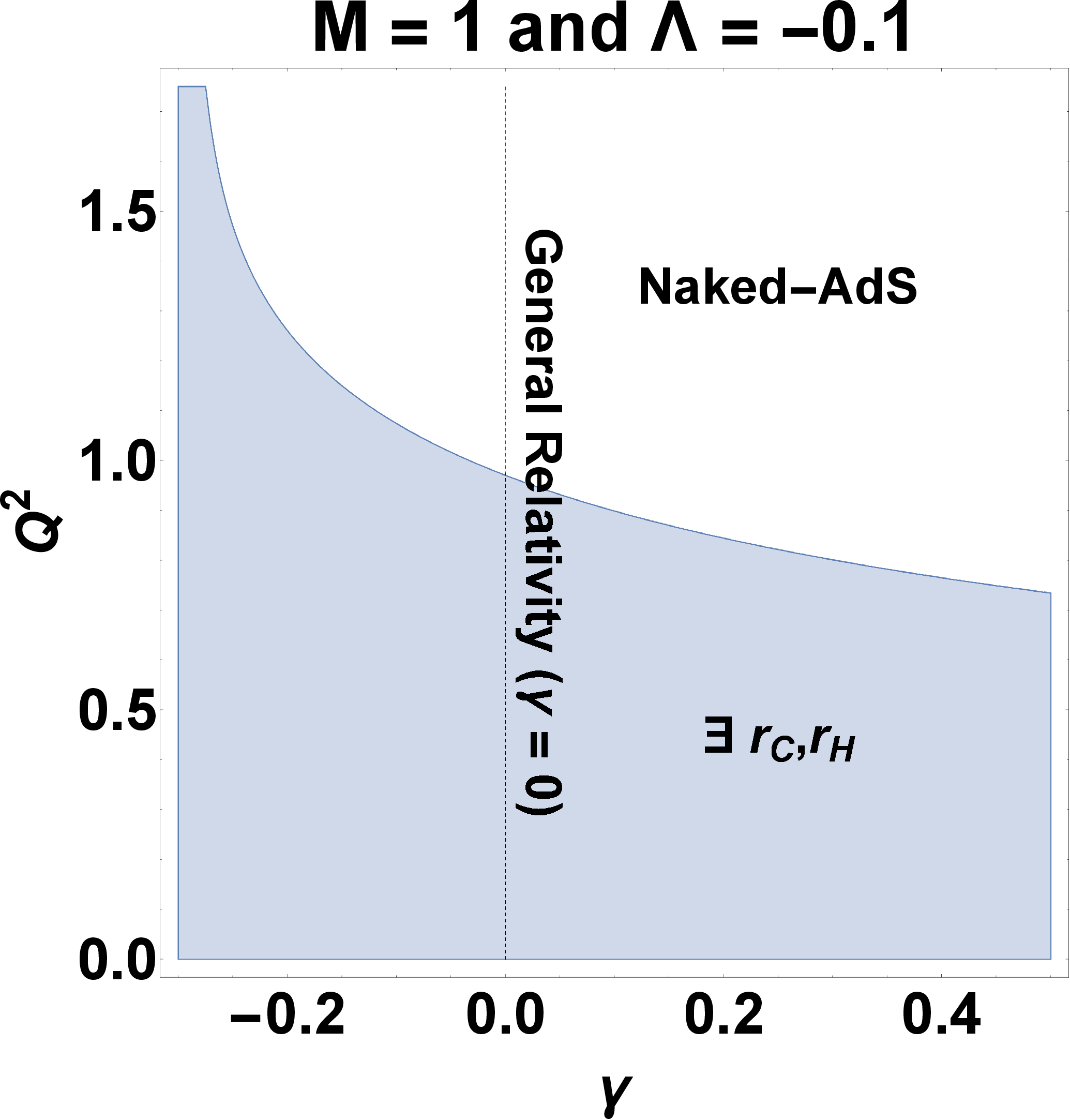}}
		\caption{Parameter space in $\gamma$-$Q^{2}$ with negative $\Lambda$.}
		\label{Sample-AdS}
	\end{figure}

The parameter space for fixed $M$ and varying $\Lambda$ are shown in FIG. \ref{Sample-AdS}.
By setting $\Lambda\rightarrow0$, the `$\exists~r_{\Lambda-},r_{\Lambda+}$' region appears on the left side of the $Q^{2}$-axis ($\gamma=0$). This region can only exist when $\gamma$ and $\Lambda$ are negative. The `$\exists~r_{C},r_{H},r_{\Lambda}$' region continues to exist in asymptotically AdS space. However, the presence of $r_{\Lambda}$ is no longer guaranteed. Hence, the region is renamed to `$\exists~r_{C},r_{H}$' in the AdS background. The overlap between these two regions is called `All Four Real Roots'. There is one other region called `No Island'. This is a region where only $r_{C}$ and $r_{\Lambda+}$ exist, that is, no `Island Universe'.

Similar to the previous case, most lines that separate each region represent loci of extremal. However, this is only true when the separated region have different numbers of real root. For example, the `$\exists~r_{\Lambda-},r_{\Lambda+}$' region has two distinct roots $r_{\Lambda-}$ and $r_{\Lambda+}$ whilst the `Naked-AdS' does not have any real root. The line between the two regions is where $r_{\Lambda-}$ and $r_{\Lambda+}$ merge into one horizon. This is a condition for extremality. In contrast, the line between `No Island' region and `$\exists~r_{\Lambda-},r_{\Lambda+}$' region does not represent any extremal locus. Both region posses the same amount of real roots. The difference between $r_{\Lambda-}$ and $r_{C}$ is only apparent when both roots exist simultaneously. In `No Island' region, $r_{C}$ is the Cauchy Horizon for the unbound Universe. In `$\exists~r_{\Lambda-},r_{\Lambda+}$' region, $r_{\Lambda-}$ act as a Cauchy Horizon for the unbound Universe beyond $r_{\Lambda+}$. The line between these regions is where $r_{C}$ becomes $r_{\Lambda-}$. However, the difference is not physically apparent.

The intersection between extremal locus represents a condition for double extremal. These conditions are marked by colored dots. The red dot represents a double extremal type $r_{C}\sim r_{H}\sim r_{\Lambda-}$. The black dot represents a double extremal type $r_{C}\sim r_{H}$ and $r_{\Lambda-}\sim r_{\Lambda+}$. The green dot represents a double extremal type $r_{H}\sim r_{\Lambda-}\sim r_{\Lambda+}$.

As the value of $\Lambda$ decreases, the `$\exists~r_{\Lambda-},r_{\Lambda+}$' region becomes smaller. At $\Lambda=-\displaystyle{\frac{1}{36}}$, the tip of the region~(green dot) locates at $\gamma=-\displaystyle{\frac{1}{6}}$ and $Q^{2}=0$. This location is also a condition for double extremal type $r_{H}\sim r_{\Lambda-}\sim r_{\Lambda+}$. As the value of $\Lambda$ decreases further, the tip slowly moving up along the extremal locus ($r_{H}\sim r_{\Lambda-}$). Once $\Lambda=-\displaystyle{\frac{1}{18}}$, the tip of `$\exists~r_{\Lambda-},r_{\Lambda+}$' region merges with the tip of the `$\exists~r_{C},r_{H}$' region. This effect causes all three dots representing double extremal to converge into one triple extremal condition $r_{C}\sim r_{H}\sim r_{\Lambda-}\sim r_{\Lambda+}$ at $\gamma=-\displaystyle{\frac{2}{9}}$ and $Q^{2}=\displaystyle{\frac{3}{2}}$. For $\Lambda<-\displaystyle{\frac{1}{18}}$, there is no longer any physical difference between `$\exists~r_{\Lambda-},r_{\Lambda+}$', `No Island' and `$\exists~r_{C},r_{H}$' region. The three regions are merged into one single region called `$\exists~r_{C},r_{H}$'.

\section{Analytic Solution of QNMs for Near-Extremal in Massive Gravity Background}\label{Math}

For neutral scalar field, we have the following equation in the generalized spherically symmetric background, $\Box\Phi(\mathbf{x})=0$. The scalar field can be expressed as $\Phi(\mathbf{r},t)=\displaystyle{\frac{\phi(r)}{r}}Y_{\ell m}(\theta,\varphi)e^{-i\omega t}$, where $Y_{\ell m}(\theta,\varphi)$ is the spherical harmonic. The radial equation then takes the following form,
	\begin{equation}\label{Initial}
		\frac{d^{2}\phi}{dr^{2}_{*}}+\left[\omega^{2}-f(r)\left(m^{2}_{s}+\frac{\ell(\ell+1)}{r^{2}}+\frac{f'(r)}{r}\right)\right]\phi=0,
	\end{equation}
where $r_{*}$ is defined by $\displaystyle{\frac{dr_{*}}{dr}=\frac{1}{f(r)}}$. Mass of the scalar is denoted by $m_{s}$. The azimuthal angular quantum number are represented by $\ell$. $\omega$ is the quasinormal frequency. The $f(r)$ is given in (\ref{gmet}). By using near-extremal limit the above equation can be simplified, the details of the calculation can be found in Ref.~\cite{Burikham:2020dfi}. The resulting equation for near-extremal limits $r_{C}\sim r_{H}$, $r_{H}\sim r_{\Lambda}$~(dS) or $r_{H}\sim r_{\Lambda-}$~(AdS) are 
	\begin{equation}
		\frac{d^{2}\phi}{dr^{2}_{*}}+\left[\omega^{2}-\frac{V_{0}}{\cosh^{2}(\kappa_{H}r_{*})}\right]\phi=0,
	\end{equation}
where
	\begin{equation}
		V_{0}=\kappa^{2}_{H}\frac{m^{2}_{s}r^{2}_{H}+\ell(\ell+1)}{1+r_{H}\gamma-2Q^{2}/r^{2}_{H}}.
	\end{equation}
For near-extremal limit $r_{\Lambda-}\sim r_{\Lambda+}$, the following result is obtained,
	\begin{equation}
		\frac{d^{2}\phi}{dr^{2}_{*}}+\left[\omega^{2}-\frac{V_{0}}{\cosh^{2}(\kappa_{\Lambda+}r_{*})}\right]\phi=0,
	\end{equation}
where
	\begin{equation}
		V_{0}=\kappa^{2}_{\Lambda+}\frac{m^{2}_{s}r^{2}_{\Lambda+}+\ell(\ell+1)}{1+r_{\Lambda+}\gamma-2Q^{2}/r^{2}_{\Lambda+}}.
	\end{equation}
These equations are differential equation with the P\"{o}schl-Teller potential. The solution has generically complex frequencies
	\begin{equation}
		\omega_{n}=\kappa_{\chi}\left\lbrace\sqrt{\frac{V_{0}}{\kappa^{2}_{\chi}}-\frac{1}{4}}-\left(n+\frac{1}{2}\right)i\right\rbrace,
	\end{equation}
where $\kappa_{\chi}$ denotes an appropriate surface gravity for the near-extremal condition \cite{Agboola:2008axa}.

\section{Numerical Analysis of QNMs of Near-Extremal Black Holes in \MakeLowercase{d}S Space}\label{dSN}

As shown previously in Section \ref{Math}, the associated quasinormal frequencies $\omega_{n}$ of extremal black hole are,
	\begin{equation}\label{QNMF}
		\omega_{n}=\kappa_{\chi}\left\lbrace\pm\sqrt{\frac{V_{0}}{\kappa^{2}_{\chi}}-\frac{1}{4}}-\left(n+\frac{1}{2}\right)i\right\rbrace,
	\end{equation}
where $\kappa_{\chi}$ is the associated surface gravity of near-extremal horizon \cite{Burikham:2020dfi,Ferrari:1984zz,Cardoso:2003sw} defined by,
	\begin{equation}
		\kappa_{\chi}=\left.\frac{1}{2}\frac{df}{dr}\right|_{r=r_{\chi}},
	\end{equation}
and $r_{\chi}$ is the extremal horizon whose surface gravity is positive. The accuracy of the above equation is governed by $\bigtriangleup r/r_{\chi}$, where $\bigtriangleup r$ is the difference between the two near-extremal horizons. The Asymptotic Iteration Method (AIM \cite{Cho:2009cj,Ponglertsakul:2018smo,Cho:2011sf}) is used to confirm the accuracy of Eqn.~(\ref{QNMF}), see Appendix~\ref{SectAIM} for more detail on AIM. We found that $\bigtriangleup r/r_{\chi}<0.02$ suffices to ensure good agreement between AIM and our formula. The value of $\bigtriangleup r$ can be reduced further by choosing a pair of $(\gamma,Q^{2})$ that is closer to extremal locus (shown in FIG. \ref{Sample-dS} and \ref{Sample-AdS}). The reduction of $\bigtriangleup r/r_{\chi}$ results in higher accuracy as well as shorter computational time.

Additionally, the Eqn.~(\ref{QNMF}) has an interesting property. If the term under the square root sign is positive, then the quasinormal frequencies have two families of frequency with positive and negative real part. However, if the term under the square root sign is negative, then the two families of frequency would have a zero real part and overlay on top of each other at the imaginary axis. The quasinormal frequencies are shown in TABLE \ref{QNMPD+}, \ref{QNMPD-} and \ref{QNMIM}.

\subsection{QNMs of Near-Extremal Black Hole with $r_{C}\sim r_{H}$}\label{RCTRH}

\begin{figure}[h]
		\centering
		\includegraphics[width=0.7\textwidth]{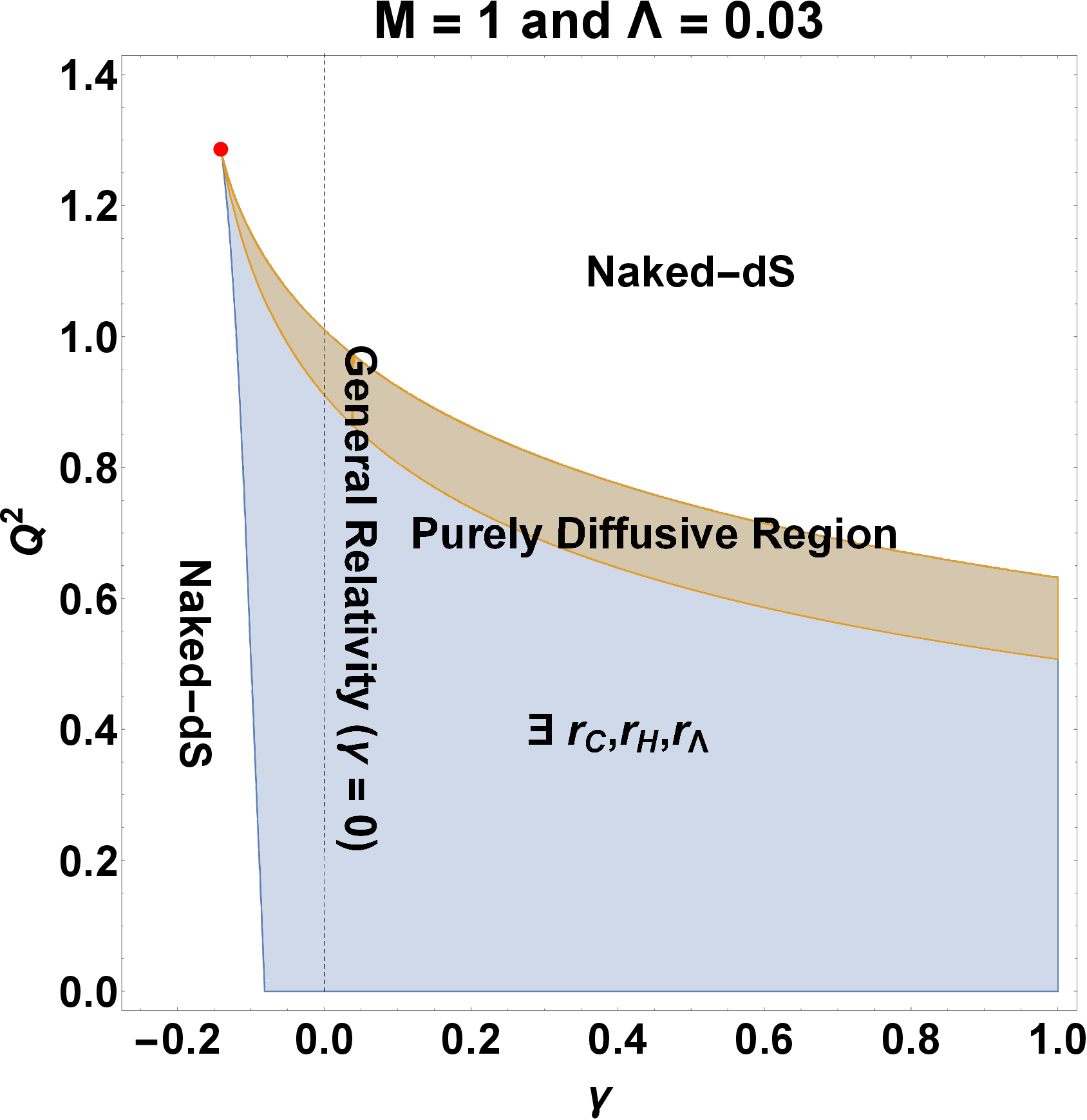}
		\caption{\raggedright{The plot of the region satisfying (\ref{RPD}) in the black hole parameter space. The condition given by (\ref{RPD}) is clearly satisfied for all loci of $r_{C}\sim r_{H}$ extremal.}}
		\label{PD}
	\end{figure}

If the term under the square root on the RHS~(right-handed side) of (\ref{QNMF}) is negative, then the quasinormal frequencies will be purely diffusive mode (zero real part). This condition requires that $V_{0}/\kappa^{2}_{\chi}\le1/4$. However, it is much more convenient to use a more restricted condition, $V_{0}/\kappa^{2}_{\chi}<0$. This condition is easier to verify and capable of guaranteeing that the quasinormal mode is a diffusive mode. The condition can be expressed as follows,
	\begin{equation}\label{RPD}
		1+r_{H}\gamma-2\left(\frac{Q}{r_{H}}\right)^{2}<0,
	\end{equation}
where $r_{H}$ is the event horizon. It can be expressed in terms of the spacetime parameters as
	\begin{equation}
		r_{H}=A+\frac{1}{2}\sqrt{\pi_{1}}-\frac{1}{2}\sqrt{\pi_{2}+\pi_{3}},\nonumber
	\end{equation}
where
	\begin{eqnarray}
		\pi_{1}&=&\left(\frac{3\gamma}{2\Lambda}\right)^{2}-\frac{1}{\Lambda}\left[\frac{1}{B^{1/3}}\big(1-4Q\Lambda+6M\gamma\big)+B^{1/3}-2\right],\nonumber \\
		\pi_{2}&=&\frac{1}{2}\left(\frac{3\gamma}{\Lambda}\right)^{2}+\frac{1}{\Lambda}\left[\frac{1}{B^{1/3}}\big(1-4Q\Lambda+6M\gamma\big)+B^{1/3}+4\right],\nonumber \\
		\pi_{3}&=&\frac{3\gamma}{\Lambda\sqrt{\pi_{1}}}\left[\left(\frac{3\gamma}{2\Lambda}\right)^{2}+\frac{3}{\Lambda}-4\frac{M}{\gamma}\right],\nonumber
	\end{eqnarray}
and
	\begin{eqnarray}
		A&=&\frac{3\gamma}{4\Lambda},\nonumber \\
		B&=&C+\sqrt{C^{2}-\big(1-4\Lambda Q+6M\gamma\big)^{3}},\nonumber \\
		C&=&1-18\Lambda M^{2}+12\Lambda Q+9M\gamma+\frac{27}{2}Q\gamma^{2}.\nonumber
	\end{eqnarray}
The region where (\ref{RPD}) is satisfied is shown in FIG. \ref{PD}. 

A sample of extremal black hole is used to confirmed the validity of the equation (\ref{QNMF}). Numerical calculations are performed by the Improved Asymptotic Iteration Method~(improved AIM). The quasinormal frequencies from the approximate analytic formula and AIM are shown in TABLE \ref{QNMPD+} and \ref{QNMPD-}.
	\begin{table}[h]
		\centering
		\begin{tabular}{|l|m{2em}|c|c|}
			\hline
			\{$M$, $Q$, $\Lambda$, $\gamma$, $\tau$\} \{$\ell$, $m_{s}$, $\kappa_{H}$, $\bigtriangleup r/r_{H}$\}&~~$n$&AIM&Formula~(\ref{QNMF})\\
			\hline
			\multirow{3}{12em}{~\{1, 0.8805, 0.03, 0.4, 1\}\\~\{0, 0.01, 0.009264, 0.007\}}&~~0&~- $4.061\times10^{-13}$ + $2.626\times10^{-7}i$~~&0 + $2.599\times10^{-7}i$\\
			&~~1&- $1.291\times10^{-7}$ - 0.00933$i$&0 - 0.00926$i$\\
			&~~2&- $1.650\times10^{-6}$ - 0.01865$i$&0 - 0.01853$i$\\
			\hline
			\multirow{3}{12em}{~\{1, 0.8805, 0.03, 0.4, 1\}\\~\{2, 0.01, 0.009264, 0.007\}}&~~0&- $2.286\times10^{-12}$ + 0.01291$i$&0 + 0.01293$i$\\
			&~~1&- $1.018\times10^{-9}$ + 0.00358$i$&0 + 0.00366$i$\\
			&~~2&- $1.428\times10^{-6}$ - 0.00574$i$&0 - 0.00560$i$\\
			\hline
			\multirow{3}{14em}{~\{1, 0.96109324, 0.03, 0.1, 1\}\\~\{0, 0.01, $5.925\times10^{-5}$, 0.00008\}}&~~0&0 - $3.8833\times10^{-9}i$&0 + $3.884\times10^{-9}i$\\
			&~~1&0 - $5.92488\times10^{-5}i$&0 - $5.92436\times10^{-5}i$\\
			&~~2&$1.4313\times10^{-9}$ - $1.18287\times10^{-4}i$&0 - $1.18491\times^{-4}i$\\
			\hline
			\multirow{3}{14em}{~\{1, 0.96109324, 0.03, 0.1, 1\}\\~\{2, 0.01, $5.925\times10^{-5}$, 0.00008\}}&~~0&0 + $1.05059\times10^{-4}i$&~0 + $1.05061\times10^{-4}i$~~\\
			&~~1&0 + $4.58073\times10^{-5}i$&0 + $4.58133\times10^{-5}i$\\
			&~~2&0 - $1.37187\times10^{-5}i$&0 - $1.34343\times10^{-5}i$\\
			\hline
		\end{tabular}
		\caption{\raggedright{The QNMs of near-extremal black hole with $r_{C}\sim r_{H}$ where $\bar{x}/x_{H}=0.7$ with a positive sign is chosen for the square root term.}}
		\label{QNMPD+}
	\end{table}
	\begin{table}[h]
		\centering
		\begin{tabular}{|l|m{2em}|c|c|}
			\hline
			\{$M$, $Q$, $\Lambda$, $\gamma$, $\tau$\} \{$\ell$, $m_{s}$, $\kappa_{H}$, $\bigtriangleup r/r_{H}$\}&~~$n$&AIM&Formula~(\ref{QNMF})\\
			\hline
			\multirow{3}{12em}{~\{1, 0.8805, 0.03, 0.4, 1\}\\~\{0, 0.01, 0.009264, 0.007\}}&~~0&$2.524\times10^{-5}$ - 0.009326$i$&0 - 0.009326$i$\\
			&~~1&$2.355\times10^{-7}$ - 0.01865$i$&0 - 0.01853$i$\\
			&~~2&$9.697\times10^{-8}$ - 0.02802$i$&0 - 0.02779$i$\\
			\hline
			\multirow{3}{12em}{~\{1, 0.8805, 0.03, 0.4, 1\}\\~\{2, 0.01, 0.009264, 0.007\}}&~~0&$1.077\times10^{-14}$ - 0.02224$i$&0 - 0.02219$i$\\
			&~~1&~- $3.241\times10^{-12}$ - 0.03156$i$~~&0 - 0.03146$i$\\
			&~~2&- $4.096\times10^{-10}$ - 0.04058$i$&0 - 0.04072$i$\\
			\hline
			\multirow{3}{14em}{~\{1, 0.96109324, 0.03, 0.1, 1\}\\~\{0, 0.01, $5.925\times10^{-5}$, 0.00008\}}&~~0&0 - $5.92563\times10^{-5}i$&~0 - $5.92514\times10^{-5}i$~~\\
			&~~1&0 - $1.18517\times10^{-4}i$&0 - $1.18499\times10^{-4}i$\\
			&~~2&0 - $1.77779\times10^{-4}i$&0 - $1.77746\times10^{-4}i$\\
			\hline
			\multirow{3}{14em}{~\{1, 0.96109324, 0.03, 0.1, 1\}\\~\{2, 0.01, $5.925\times10^{-5}$, 0.00008\}}&~~0&0 - $1.64311\times10^{-4}i$&0 - $1.64308\times10^{-4}i$\\
			&~~1&0 - $2.23564\times10^{-4}i$&0 - $2.23556\times10^{-4}i$\\
			&~~2&0 - $2.82817\times10^{-4}i$&0 - $2.82803\times10^{-4}i$\\
			\hline
		\end{tabular}
		\caption{\raggedright{The QNMs of near-extremal black hole for negative sign of the square root term.}}
		\label{QNMPD-}
	\end{table}

\subsection{QNMs of Near-Extremal Black Hole with $r_{H}\sim r_{\Lambda}$}\label{RHTRLam}

For this case, it is possible to have QNMs with non-zero real part. This can be achieved by setting $\ell\ge2$. The improved AIM~(see Appendix~\ref{SectAIM}) is used with $n=250$ to perform the numerical calculation. The physical Universe is extremely small ($x_{H}-x_{\Lambda}\ll1$, where $x_{H,\Lambda}=1/r_{H,\Lambda}$ respectively). We thus choose $\bar{x}=2x_{H}x_{\Lambda}/(x_{H}+x_{\Lambda})$ in order to find the converging solution. The results are shown in TABLE \ref{QNMIM}.
	\begin{table}[h]
		\centering
		\begin{tabular}{|c|m{2em}|c|c|}
			\hline
			\{$M$, $Q$, $\Lambda$, $\gamma$, $\tau$\} \{$\ell$, $m_{s}$, $\kappa_{H}$, $\bigtriangleup r/r_{H}$\}~&~~$n$&AIM~($10^{-5}$)&Formula~(\ref{QNMF})~($10^{-5}$)\\
			\hline
			\multirow{3}{14em}{\{1, 0.1, 0.03, -0.08771526, 1\}\\\{2, 0.01, $3.581\times10^{-5}$, 0.00037\}}&~~0&~$\pm10.426$ - 1.790$i$~~&~$\pm10.427$ - 1.790$i$~~\\
			&~~1&$\pm10.426$ - 5.370$i$&$\pm10.427$ - 5.371$i$\\
			&~~2&$\pm10.413$ - 8.947$i$&$\pm10.427$ - 8.952$i$\\
			\hline
			\multirow{3}{14em}{\{1, 0.2, 0.03, -0.08839355, 1\}\\\{2, 0.01, $2.471\times10^{-5}$, 0.00025\}}&~~0&$\pm7.225$ - 1.2352$i$&$\pm7.225$ - 1.2353$i$\\
			&~~1&$\pm7.225$ - 3.7057$i$&$\pm7.225$ - 3.7060$i$\\
			&~~2&$\pm7.225$ - 6.1761$i$&$\pm7.225$ - 6.1767$i$\\
			\hline
			\multirow{3}{14em}{\{1, 0.5, 0.09, -0.03040959, 1\}\\\{2, 0.01, $2.617\times10^{-5}$, 0.00018\}}&~~0&$\pm6.815$ - 1.3082$i$&$\pm6.816$ - 1.3083$i$\\
			&~~1&$\pm6.815$ - 3.9242$i$&$\pm6.816$ - 3.9248$i$\\
			&~~2&$\pm6.808$ - 6.5036$i$&$\pm6.816$ - 6.5413$i$\\
			\hline
		\end{tabular}
		\caption{The QNMs of near-extremal black hole with $r_{H}\sim r_{\Lambda}$ in units of $10^{-5}$.  }
		\label{QNMIM}
	\end{table}

\section{Numerical Analysis of QNMs of Near-Extremal Black Hole in A\MakeLowercase{d}S Space}\label{AdSN}

For the asymptotically AdS, there are three different types of near-extremal black hole. Some of which contains near-extremal horizon as the largest horizon. For these cases, the AIM cannot be applied due to the different boundary conditions of the waves. The method requires that the Physical Universe is bounded by horizon on both sides. In such situation, the Spectral Method is used to calculate the QNMs with vanishing profile at the AdS boundary, see Appendix~\ref{SM} for details. On the other hand for the ``Island Universe'' in the asymptotically AdS, the characteristic of $f(r)$ as well as the boundary conditions of the scalar field are locally similar to the asymptotically dS case in the region $r_{H}<r<r_{\Lambda-}$. The improved AIM can be used in this case.

The QNMs of the near-extremal black hole still obey the same formula (\ref{QNMF})~(see Ref.~\cite{Burikham:2020dfi} for details).

\subsection{QNMs of Near-Extremal Black Hole with $r_{C}\sim r_{H}$}

From FIG. \ref{Sample-AdS}, the horizontal curve represents the extremal locus with $r_{C}\sim r_{H}$. Along this curve, the AIM can only be used for the points overlapping with `$\exists~r_{\Lambda-},r_{\Lambda+}$' region. These points correspond to extremal black holes with $r_{C}\sim r_{H}$ in the presence of $r_{\Lambda-}$ and $r_{\Lambda+}$. For other points, the Spectral Method is otherwise used to calculate the QNMs of near-extremal black hole whose $r_{\Lambda-}$ and $r_{\Lambda+}$ are absent.

\subsubsection{QNMs of Near-Extremal Black Hole with $r_{C}\sim r_{H}$ in the presence of $r_{\Lambda+}$ and $r_{\Lambda-}$}

We are interested in the physical Universe between $r_{H}$ and $r_{\Lambda-}$. This Universe is similar to the physical Universe in Section \ref{RCTRH}. AIM can be used directly to calculate the QNMs. While there is another physical Universe beyond $r_{\Lambda+}$. This physical Universe is not adjacent to any near-extremal horizon. It is irrelevant to our consideration.

The calculated QNMs are shown in TABLE \ref{QNMPDA+} and \ref{QNMPDA-}.
	\begin{table}[h]
		\centering
		\begin{tabular}{|c|m{2em}|c|c|}
			\hline
			\{$M$, $Q$, $\Lambda$, $\gamma$, $\tau$\} \{$\ell$, $m_{s}$, $\kappa_{H}$, $\bigtriangleup r/r_{H}$\}&~~$n$&AIM&Formula~(\ref{QNMF})\\
			\hline
			\multirow{3}{14em}{\{1, 1.11032641, -0.015, -0.15, 1\}\\\{0, 0.01, $2.452\times10^{-5}$, 0.00017\}}&~~0&~- $6.032\times10^{-21}$ + $1.205\times10^{-8}i$~~&~0 + $1.206\times10^{-8}i$~~\\
			&~~1&$8.094\times10^{-14}$ - $2.451\times10^{-5}i$&0 - $2.451\times10^{-5}i$\\
			&~~2&- $7.776\times10^{-11}$ - $4.903\times10^{-5}i$&0 - $4.902\times10^{-5}i$\\
			\hline
			\multirow{3}{14em}{\{1, 1.11032641, -0.015, -0.15, 1\}\\\{1, 0.01, $2.452\times10^{-5}$, 0.00017\}}&~~0&- $1.966\times10^{-16}$ + $4.282\times10^{-5}i$&0 + $4.283\times10^{-5}i$\\
			&~~1&- $1.968\times10^{-14}$ + $1.830\times10^{-5}i$&0 + $1.831\times10^{-5}i$\\
			&~~2&- $9.018\times10^{-12}$ - $6.228\times10^{-6}i$&0 - $6.211\times10^{-6}i$\\
			\hline
			\multirow{3}{14em}{\{1, 1.11032641, -0.015, -0.15, 1\}\\\{2, 0.01, $2.452\times10^{-5}$, 0.00017\}}&~~0&- $2.181\times10^{-16}$ + $8.156\times10^{-5}i$&0 + $8.156\times10^{-5}i$\\
			&~~1&$1.008\times10^{-13}$ + $5.703\times10^{-5}i$&0 + $5.704\times10^{-5}i$\\
			&~~2&- $4.792\times10^{-12}$ + $3.251\times10^{-5}i$&0 + $3.252\times10^{-5}i$\\
			\hline
			\multirow{3}{14em}{\{1, 1.0275949, -0.0015, -0.05, 1\}\\\{0, 0.01, $2.139\times10^{-4}$, 0.00055\}}&~~0&- $1.937\times10^{-20}$ + $3.010\times10^{-8}i$&0 + $3.014\times10^{-8}i$\\
			&~~1&$1.016\times10^{-12}$ - $2.139\times10^{-4}i$&0 - $2.138\times10^{-4}i$\\
			&~~2&- $2.135\times10^{-9}$ - $4.278\times10^{-4}i$&0 - $4.277\times10^{-4}i$\\
			\hline
			\multirow{3}{14em}{\{1, 1.0275949, -0.0015, -0.05, 1\}\\\{1, 0.01, $2.139\times10^{-4}$, 0.00055\}}&~~0&- $9.142\times10^{-17}$ + $2.400\times10^{-4}i$&0 + $2.401\times10^{-4}i$\\
			&~~1&- $7.400\times10^{-14}$ + $2.604\times10^{-5}i$&0 + $2.621\times10^{-5}i$\\
			&~~2&- $3.952\times10^{-10}$ - $1.879\times10^{-4}i$&0 - $1.877\times10^{-4}i$\\
			\hline
			\multirow{3}{14em}{\{1, 1.0275949, -0.0015, -0.05, 1\}\\\{2, 0.01, $2.139\times10^{-4}$, 0.00055\}}&~~0&- $1.777\times10^{-16}$ + $4.747\times10^{-4}i$&0 + $4.747\times10^{-4}i$\\
			&~~1&$7.800\times10^{-14}$ + $2.607\times10^{-4}i$&0 + $2.609\times10^{-4}i$\\
			&~~2&$2.350\times10^{-11}$ + $4.675\times10^{-5}i$&0 + $4.702\times10^{-5}i$\\
			\hline
		\end{tabular}
		\caption{\raggedright{The QNMs of near-extremal black hole with $r_{C}\sim r_{H}$ in the presence of $r_{\Lambda-}$ and $r_{\Lambda+}$. The $\bar{x}$ is set to $\bar{x}/x_{H}=0.8$. The sign of the square root term of (\ref{QNMF}) is chosen to be positive.}}
		\label{QNMPDA+}
	\end{table}
	\begin{table}[h]
		\centering
		\begin{tabular}{|c|m{2em}|c|c|}
			\hline
			\{$M$, $Q$, $\Lambda$, $\gamma$, $\tau$\} \{$\ell$, $m_{s}$, $\kappa_{H}$, $\bigtriangleup r/r_{H}$\}&~~$n$&AIM&Formula~(\ref{QNMF})\\
			\hline
			\multirow{3}{14em}{\{1, 1.11032641, -0.015, -0.15, 1\}\\\{0, 0.01, $2.452\times10^{-5}$, 0.00017\}}&~~0&$4.738\times10^{-14}$ - $2.454\times10^{-5}i$&0 - $2.453\times10^{-5}i$\\
			&~~1&~- $1.423\times10^{-10}$ - $4.906\times10^{-5}i$~~&~0 - $4.905\times10^{-5}i$~~\\
			&~~2&$2.697\times10^{-12}$ - $7.356\times10^{-5}i$&0 - $7.357\times10^{-5}i$\\
			\hline
			\multirow{3}{14em}{\{1, 1.11032641, -0.015, -0.15, 1\}\\\{1, 0.01, $2.452\times10^{-5}$, 0.00017\}}&~~0&$7.193\times10^{-18}$ - $6.735\times10^{-5}i$&0 - $6.734\times10^{-5}i$\\
			&~~1&- $9.306\times10^{-16}$ - $9.187\times10^{-5}i$&0 - $9.186\times10^{-5}i$\\
			&~~2&- $7.495\times10^{-13}$ - $1.164\times10^{-4}i$&0 - $1.164\times10^{-4}i$\\
			\hline
			\multirow{3}{14em}{\{1, 1.11032641, -0.015, -0.15, 1\}\\\{2, 0.01, $2.452\times10^{-5}$, 0.00017\}}&~~0&$1.850\times10^{-17}$ - $1.061\times10^{-4}i$&0 - $1.061\times10^{-4}i$\\
			&~~1&- $9.882\times10^{-16}$ - $1.306\times10^{-4}i$&0 - $1.306\times10^{-4}i$\\
			&~~2&- $1.290\times10^{-12}$ - $1.551\times10^{-4}i$&0 - $1.551\times10^{-4}i$\\
			\hline
			\multirow{3}{14em}{\{1, 1.0275949, -0.0015, -0.05, 1\}\\\{0, 0.01, $2.139\times10^{-4}$, 0.00055\}}&~~0&$1.016\times10^{-12}$ - $2.139\times10^{-4}i$&0 - $2.1382\times10^{-4}i$\\
			&~~1&- $2.135\times10^{-9}$ - $4.278\times10^{-4}i$&0 - $4.278\times10^{-4}i$\\
			&~~2&- $1.824\times10^{-11}$ - $6.419\times10^{-4}i$&0 - $6.416\times10^{-4}i$\\
			\hline
			\multirow{3}{14em}{\{1, 1.0275949, -0.0015, -0.05, 1\}\\\{1, 0.01, $2.139\times10^{-4}$, 0.00055\}}&~~0&$1.418\times10^{-14}$ - $4.540\times10^{-4}i$&0 - $4.539\times10^{-4}i$\\
			&~~1&$1.093\times10^{-13}$ - $6.680\times10^{-4}i$&0 - $6.678\times10^{-4}i$\\
			&~~2&- $9.918\times10^{-12}$ - $8.820\times10^{-4}i$&0 - $8.817\times10^{-4}i$\\
			\hline
			\multirow{3}{14em}{\{1, 1.0275949, -0.0015, -0.05, 1\}\\\{2, 0.01, $2.139\times10^{-4}$, 0.00055\}}&~~0&$1.262\times10^{-16}$ - $6.887\times10^{-4}i$&0 - $6.886\times10^{-4}i$\\
			&~~1&- $6.606\times10^{-16}$ - $9.027\times10^{-4}i$&0 - $9.025\times10^{-4}i$\\
			&~~2&- $1.705\times10^{-11}$ - $1.117\times10^{-3}i$&0 - $1.116\times10^{-3}i$\\
			\hline
		\end{tabular}
		\caption{\raggedright{The QNMs of near-extremal black hole with $r_{C}\sim r_{H}$ in the presence of $r_{\Lambda-}$ and $r_{\Lambda+}$. $\bar{x}/x_{H}$ is set to $0.8$.}}
		\label{QNMPDA-}
	\end{table}

\subsubsection{QNMs of Near-Extremal Black Hole with $r_{C}\sim r_{H}$ where $r_{\Lambda+}$ and $r_{\Lambda-}$ Do Not Existed}\label{RCTRHwoRLam}

The value of $r$ in the spacetime of interest has the range of $[r_{H},\infty)$. This unbounded nature requires a modification to the AIM. We instead adopt the Spectral Method for the numerical calculation of the QNMs in this case. The results obtained from this method only consists of a single family of $\omega_{n}$ namely,
	\begin{equation}
			\omega_{n}=\kappa_{H}\left\lbrace-\sqrt{\frac{V_{0}}{\kappa^{2}_{H}}-\frac{1}{4}}-\left(n+\frac{1}{2}\right)i\right\rbrace.
	\end{equation}
The Spectral Method is much less time consuming compared to the AIM. More details of the method can be found in Appendix~\ref{SM} and references therein.

The results obtained from the Spectral Method are shown in TABLE \ref{QNMCMT}. Note that the results from the Spectral Method are given in terms of $w_{n}=\omega_{n} r_{H}$.
	\begin{table}[h]
		\centering
		\begin{tabular}{|c|m{2em}|c|c|}
			\hline
			\{$M$, $Q$, $\Lambda$, $\gamma$, $\tau$\} \{$\ell$, $m_{s}$, $\kappa_{H}$, $\bigtriangleup r/r_{H}$\}&~~$n$&Spectral Method&Formula~(\ref{QNMF})\\
			\hline
			\multirow{3}{14em}{\{1, 0.8988057, -0.03, 0.3, 1\}\\\{0, 0.01, $7.197\times10^{-4}$, 0.00063\}}&~~0&~0 - $5.355\times10^{-4}i$~~&~0 - $5.351\times10^{-4}i$~~\\
			&~~1&0 - 0.001071$i$&0 - 0.001070$i$\\
			&~~2&0 - 0.001606$i$&0 - 0.001605$i$\\
			\hline
			\multirow{3}{14em}{\{1, 0.8988057, -0.03, 0.3, 1\} \{1, 0.01, $7.197\times10^{-4}$, 0.00063\}}&~~0&0 - $9.069\times10^{-4}i$&0 - $9.066\times10^{-4}i$\\
			&~~1&0 - 0.001442$i$&0 - 0.001442$i$\\
			&~~2&0 - 0.001979$i$&0 - 0.001977$i$\\
			\hline
			\multirow{3}{14em}{\{1, 0.8988057, -0.03, 0.3, 1\} \{2, 0.01, $7.197\times10^{-4}$, 0.00063\}}&~~0&0 - 0.001308$i$&0 - 0.001308$i$\\
			&~~1&0 - 0.001844$i$&0 - 0.001843$i$\\
			&~~2&0 - 0.002389$i$&0 - 0.002378$i$\\
			\hline
			\multirow{3}{14em}{\{1, 0.9547402, -0.03, 0.1, 1\} \{0, 0.01, $6.411\times10^{-4}$, 0.00086\}}&~~0&0 - $5.601\times10^{-4}i$&0 - $5.596\times10^{-4}i$\\
			&~~1&0 - 0.001120$i$&0 - 0.001119$i$\\
			&~~2&0 - 0.001680$i$&0 - 0.001679$i$\\
			\hline
			\multirow{3}{14em}{\{1, 0.9547402, -0.03, 0.1, 1\} \{1, 0.01, $6.411\times10^{-4}$, 0.00086\}}&~~0&0 - 0.001027$i$&0 - 0.001027$i$\\
			&~~1&0 - 0.001587$i$&0 - 0.001587$i$\\
			&~~2&0 - 0.002148$i$&0 - 0.002146$i$\\
			\hline
			\multirow{3}{14em}{\{1, 0.9547402, -0.03, 0.1, 1\} \{2, 0.01, $6.411\times10^{-4}$, 0.00086\}}&~~0&0 - 0.001512$i$&0 - 0.001512$i$\\
			&~~1&0 - 0.002072&0 - 0.002071$i$\\
			&~~2&0 - 0.002635&0 - 0.002631$i$\\
			\hline
		\end{tabular}
		\caption{\raggedright{The QNMs of near-extremal black hole with $r_{C}\sim r_{H}$. Both $r_{\Lambda-}$ and $r_{\Lambda+}$ are absent.}}
		\label{QNMCMT}
	\end{table}

\subsection{QNMs of Near-Extremal Black Hole with $r_{H}\sim r_{\Lambda-}$}

There is a tiny physical Universe between $r_{H}$ and $r_{\Lambda-}$. The QNMs originated from this Universe is similar to the frequencies found in Section \ref{RHTRLam}. The $\bar{x}$ is set to $\bar{x}=2x_{H}x_{\Lambda-}/(x_{H}+x_{\Lambda-})$. The numerical results again are consistent with the approximate formula (\ref{QNMF}). Our results are shown in TABLE \ref{QNMIMA}.
	\begin{table}[h]
		\centering
		\begin{tabular}{|c|m{2em}|c|c|}
			\hline
			\{$M$, $Q$, $\Lambda$, $\gamma$, $\tau$\} \{$\ell$, $m_{s}$, $\kappa_{H}$, $\bigtriangleup r/r_{H}$\}~&~~$n$&AIM~($10^{-5}$)&Formula~(\ref{QNMF})~($10^{-5}$)\\
			\hline
			\multirow{3}{14em}{\{1, 0.5, -0.015, -0.1490915, 1\}\\\{2, 0.01, $7.310\times10^{-5}$, 0.0018\}}&~~0&$~\pm30.14$ - $3.651i$~~&~$\pm30.15$ - $3.655i$~~\\
			&~~1&$\pm30.14$ - $10.95i$&$\pm30.15$ - $10.96i$\\
			&~~2&$\pm30.14$ - $18.25i$&$\pm30.15$ - $18.27i$\\
			\hline
			\multirow{3}{14em}{\{1, 0.3, -0.015, -0.147034169, 1\}\\\{2, 0.01, $6.186\times10^{-6}$, 0.00015\}}&~~0&$\pm2.539$ - $0.3093i$&$\pm2.539$ - $0.3093i$\\
			&~~1&$\pm2.539$ - $0.9278i$&$\pm2.539$ - $0.9279i$\\
			&~~2&$\pm2.537$ - $1.547i$&$\pm2.539$ - $1.546i$\\
			\hline
			\multirow{3}{14em}{\{1, 0.8, -0.015, -0.15533078, 1\}\\\{2, 0.01, $1.075\times10^{-5}$, 0.00025\}}&~~0&$\pm4.580$ - $0.5376i$&$\pm4.581$ - $0.5377i$\\
			&~~1&$\pm4.580$ - $1.613i$&$\pm4.581$ - $1.613i$\\
			&~~2&$\pm4.580$ - $2.689i$&$\pm4.581$ - $2.688i$\\
			\hline
		\end{tabular}
		\caption{The QNMs of near-extremal black hole with $r_{H}\sim r_{\Lambda-}$ in units of $10^{-5}$. }
		\label{QNMIMA}
	\end{table}

\subsection{QNMs of Near-Extremal Black Hole with $r_{\Lambda-}\sim r_{\Lambda+}$}

Similar to the situation in Section \ref{RCTRHwoRLam}, the range of $r$ for the physical Universe is unbounded. The Spectral Method is deployed with the result only covers the solution of the negative sign of the square root in the RHS of (\ref{QNMF}). Despite having quite different $f(r)$ from Section \ref{RCTRHwoRLam}; at the region between the outer extremal horizon and beyond, the $f(r)$ behaves sufficiently similar to the previous case that the numerical calculation only requires minor adjustment. The results are shown in TABLE \ref{QNMCMH}. It is worth reminding that the results are given in $w_{n}=\omega_{n} r_{\Lambda+}$, instead of $\omega_{n}$.
	\begin{table}[h]
		\centering
		\begin{tabular}{|c|m{2em}|c|c|}
			\hline
			\{$M$, $Q$, $\Lambda$, $\gamma$, $\tau$\} \{$\ell$, $m_{s}$, $\kappa_{\Lambda+}$, $\bigtriangleup r/r_{\Lambda+}$\}&~~$n$&~Spectral Method~($10^{-4}$)~~&~Formula~(\ref{QNMF})~($10^{-4}$)~~\\
			\hline
			\multirow{3}{14em}{\{1, 0.75, -0.015, -0.12970292, 1\}\\\{0, 0.01, $1.595\times10^{-5}$, 0.00073\}}&~~0&0 - 1.843$i$&0 - 1.889$i$\\
			&~~1&0 - 3.686$i$&0 - 3.730$i$\\
			&~~2&0 - 5.529$i$&0 - 5.572$i$\\
			\hline
			\multirow{3}{14em}{\{1, 0.75, -0.015, -0.12970292, 1\}\\\{1, 0.01, $1.595\times10^{-5}$, 0.00073\}}&~~0&0 - 4.697$i$&0 - 4.707$i$\\
			&~~1&0 - 6.539$i$&0 - 6.549$i$\\
			&~~2&0 - 8.397$i$&0 - 8.390$i$\\
			\hline
			\multirow{3}{14em}{\{1, 0.75, -0.015, -0.12970292, 1\}\\\{2, 0.01, $1.595\times10^{-5}$, 0.00073\}}&~~0&0 - 7.331$i$&0 - 7.317$i$\\
			&~~1&0 - 9.194$i$&0 - 9.153$i$\\
			&~~2&0 - 11.256$i$&0 - 10.990$i$\\
			\hline
			\multirow{3}{14em}{\{1, 0.85, -0.015, -0.12980642, 1\}\\\{0, 0.01, $1.122\times10^{-5}$, 0.00051\}}&~~0&0 - 1.301$i$&0 - 1.333$i$\\
			&~~1&0 - 2.602$i$&0 - 2.633$i$\\
			&~~2&0 - 3.904$i$&0 - 3.933$i$\\
			\hline
			\multirow{3}{14em}{\{1, 0.85, -0.015, -0.12980642, 1\}\\\{1, 0.01, $1.122\times10^{-5}$, 0.00051\}}&~~0&0 - 3.295$i$&0 - 3.302$i$\\
			&~~1&0 - 4.599$i$&0 - 4.602$i$\\
			&~~2&0 - 5.982$i$&0 - 5.902$i$\\
			\hline
			\multirow{3}{14em}{\{1, 0.85, -0.015, -0.12980642, 1\}\\\{2, 0.01, $1.122\times10^{-5}$, 0.00051\}}&~~0&0 - 5.138$i$&0 - 5.141$i$\\
			&~~1&0 - 6.436$i$&0 - 6.441$i$\\
			&~~2&0 - 7.767$i$&0 - 7.741$i$\\
			\hline
		\end{tabular}
		\caption{The QNMs of near-extremal black hole with $r_{\Lambda-}\sim r_{\Lambda+}$ in units of $10^{-4}$.}
		\label{QNMCMH}
	\end{table}

\section{Black String} \label{BS}

\subsection{QNMs of Near-Extremal Black String in asymptotically dS}

Physically speaking, a black string is very different from a black hole. However, the equations of motion of the scalar field in the black string background takes similar form to the scalar field in the black hole background in certain coordinates where the symmetry~(cylindrical and spherical symmetry for the black string and black hole respectively) is manifest~\cite{Burikham:2020dfi}. The metric of the dRGT black string is given by \cite{Tannukij:2017jtn},
	\begin{equation}
		ds^{2}=-f(r)dt^{2}+\frac{dr^{2}}{f(r)}+r^{2}\big(\alpha^{2}_{g}dz^{2}+d\varphi^{2}\big), \notag
	\end{equation}
where
	\begin{equation}
		f(r)=\alpha^{2}_{m}r^{2}-\frac{2M}{r}+\frac{Q^{2}}{r^{2}}+\gamma r+\epsilon_{0}.\notag
	\end{equation}
The Klein-Gordon equation in radial direction takes the form~ \cite{Ponglertsakul:2018smo},
	\begin{equation}
		\frac{d^{2}\phi}{dr^{2}_{*}}+\left[\omega^{2}-f\left( m^{2}_{s}+\frac{1}{r^{2}}\left(\lambda^{2}+\frac{k^{2}}{\alpha^{2}_{g}}\right)+\frac{f'}{r}\right)\right]\phi=0, \notag
	\end{equation}
where $\omega,k$ and $\lambda$ are the frequency, the wave number and angular quantum number of the scalar perturbation respectively.

AIM is again applicable and only requires small modification in the effective potential and wave equation. The QNMs formula is,
	\begin{equation}\label{bseq}
		\omega_{n}=\kappa_{\chi}\left\lbrace\pm\sqrt{\frac{V_{0}}{\kappa^{2}_{\chi}}-\frac{1}{4}}-\left(n+\frac{1}{2}\right)i\right\rbrace,
	\end{equation}
where
	\begin{equation}
		V_{0}=\frac{\kappa^{2}_{\chi}}{(r_{\chi}\gamma+\epsilon-2Q^{2}/r^{2}_{\chi})}\left[m^{2}_{s}r^{2}_{\chi}+\lambda^{2}+\left(\frac{k}{\alpha_{g}}\right)^{2}\right].
	\end{equation}
$\chi$ index is used to represent extremal horizon whose surface gravity is positive. All other details of calculation from black hole cases are applied. The results are shown in TABLE \ref{QNMPDS+}, \ref{QNMPDS-} and \ref{QNMIMS}.
	\begin{table}[h]
		\centering
		\begin{tabular}{|c|m{2em}|c|c|}
			\hline
			\{$M$, $Q$, $\alpha^{2}_{m}$, $\gamma$, $\epsilon$\} \{$\lambda$, $m_{s}$, $\kappa_{H}$, $\bigtriangleup r/r_{H}$\}&~~$n$&AIM&Formula~(\ref{bseq})\\
			\hline
			\multirow{3}{13em}{\{1, 0.8805067, -0.01, 0.4, 1\}\\\{0, 0.01, 0.001072, 0.00083\}}&~~0&~- $3.152\times10^{-16}$ + $4.230\times10^{-4}i$~~&~0 + $4.231\times10^{-4}i$~~\\
			&~~1&- $1.262\times10^{-14}$ - $6.495\times10^{-4}i$&0 - $6.485\times10^{-4}i$\\
			&~~2&$1.859\times10^{-11}$ - 0.001721$i$&0 - 0.001720$i$\\
			\hline
			\multirow{3}{13em}{\{1, 0.8805067, -0.01, 0.4, 1\}\\\{1, 0.01, 0.001072, 0.00083\}}&~~0&- $7.954\times10^{-16}$ + $7.098\times10^{-4}i$&0 + $7.100\times10^{-4}i$\\
			&~~1&- $4.735\times10^{-13}$ - $3.627\times10^{-4}i$&0 - $3.616\times10^{-4}i$\\
			&~~2&$5.875\times10^{-10}$ - 0.001436$i$&0 - 0.001433$i$\\
			\hline
			\multirow{3}{13em}{\{1, 0.8805067, -0.01, 0.4, 1\}\\\{2, 0.01, 0.001072, 0.00083\}}&~~0&- $3.394\times10^{-16}$ + 0.001321$i$&0 + 0.001321$i$\\
			&~~1&- $8.159\times10^{-17}$ + $2.487\times10^{-4}i$&0 + $2.498\times10^{-4}i$\\
			&~~2&$7.177\times10^{-13}$ - $8.236\times10^{-4}i$&0 - $8.218\times10^{-4}i$\\
			\hline
			\multirow{3}{14em}{\{1, 0.9879324, -0.01, 0.035, 1\} \{0, 0.01, $4.837\times10^{-4}$, 0.0009\}}&~~0&- $1.164\times10^{-16}$ + $2.897\times10^{-4}i$&0 + $2.898\times10^{-4}i$\\
			&~~1&- $4.756\times10^{-15}$ - $1.944\times10^{-4}i$&0 - $1.938\times10^{-4}i$\\
			&~~2&$9.421\times10^{-12}$ - $6.785\times10^{-4}i$&0 - $6.775\times10^{-4}i$\\
			\hline
			\multirow{3}{14em}{\{1, 0.9879324, -0.01, 0.035, 1\} \{1, 0.01, $4.837\times10^{-4}$, 0.0009\}}&~~0&- $5.314\times10^{-17}$ + $4.700\times10^{-4}i$&0 + $4.701\times10^{-4}i$\\
			&~~1&- $2.534\times10^{-14}$ - $1.414\times10^{-5}i$&0 - $1.356\times10^{-5}i$\\
			&~~2&$2.866\times10^{-10}$ - $4.984\times10^{-4}i$&0 - $4.972\times10^{-4}i$\\
			\hline
			\multirow{3}{14em}{\{1, 0.9879324, -0.01, 0.035, 1\} \{2, 0.01, $4.837\times10^{-4}$, 0.0009\}}&~~0&$1.254\times10^{-15}$ + $8.440\times10^{-4}i$&0 + $8.442\times10^{-4}i$\\
			&~~1&$5.178\times10^{-14}$ + $3.599\times10^{-4}i$&0 + $3.605\times10^{-4}i$\\
			&~~2&$1.806\times10^{-11}$ - $1.241\times10^{-4}i$&0 - $1.232\times10^{-4}i$\\
			\hline
		\end{tabular}
		\caption{\raggedright{The QNMs of near-extremal black string. The value of $Q$ is fine tuned such that $r_{C}\sim r_{H}$. The sign for square root term is chosen to be positive. The value of $k$ and $\alpha_{g}$ is chosen to be 1. $\bar{x}/x_{H}$ is set to 0.8.}}
		\label{QNMPDS+}
	\end{table}
	\begin{table}[h]
		\centering
		\begin{tabular}{|c|m{2em}|c|c|}
			\hline
			\{$M$, $Q$, $\alpha^{2}_{m}$, $\gamma$, $\epsilon$\} \{$\lambda$, $m_{s}$, $\kappa_{H}$, $\bigtriangleup r/r_{H}$\}&~~$n$&AIM&Formula~(\ref{bseq})\\
			\hline
			\multirow{3}{13em}{\{1, 0.8805067, -0.01, 0.4, 1\}\\\{0, 0.01, 0.001072, 0.00083\}}&~~0&- $5.536\times10^{-15}$ - 0.001495$i$&0 - 0.001495$i$\\
			&~~1&- $1.729\times10^{-12}$ - 0.002568$i$&0 - 0.002566$i$\\
			&~~2&- $1.281\times10^{-10}$ - 0.003637$i$&0 - 0.003638$i$\\
			\hline
			\multirow{3}{13em}{\{1, 0.8805067, -0.01, 0.4, 1\}\\\{1, 0.01, 0.001072, 0.00083\}}&~~0&- $3.942\times10^{-15}$ - 0.001782$i$&0 - 0.001782$i$\\
			&~~1&$1.272\times10^{-14}$ - 0.002855$i$&0 - 0.002853$i$\\
			&~~2&- $1.062\times10^{-11}$ - 0.003927$i$&0 - 0.003925$i$\\
			\hline
			\multirow{3}{13em}{\{1, 0.8805067, -0.01, 0.4, 1\}\\\{2, 0.01, 0.001072, 0.00083\}}&~~0&$8.581\times10^{-16}$ - 0.002394$i$&0 - 0.002393$i$\\
			&~~1&$4.673\times10^{-14}$ - 0.003466$i$&0 - 0.003465$i$\\
			&~~2&$3.372\times10^{-13}$ - 0.004540$i$&0 - 0.004536$i$\\
			\hline
			\multirow{3}{14em}{\{1, 0.9879324, -0.01, 0.035, 1\}\\\{0, 0.01, $4.837\times10^{-4}$, 0.0009\}}&~~0&~- $1.523\times10^{-15}$ - $7.738\times10^{-4}i$~~&~0 - $7.735\times10^{-4}i$~~\\
			&~~1&$6.658\times10^{-14}$ - 0.001258$i$&0 - 0.001257$i$\\
			&~~2&- $1.472\times10^{-12}$ - 0.001742$i$&0 - 0.001741$i$\\
			\hline
			\multirow{3}{14em}{\{1, 0.9879324, -0.01, 0.035, 1\}\\\{1, 0.01, $4.837\times10^{-4}$, 0.0009\}}&~~0&- $2.912\times10^{-16}$ - $9.541\times10^{-4}i$&0 - $9.538\times10^{-4}i$\\
			&~~1&$1.135\times10^{-14}$ - 0.001438$i$&0 - 0.001438$i$\\
			&~~2&- $3.792\times10^{-11}$ - 0.001921$i$&0 - 0.001921$i$\\
			\hline
			\multirow{3}{14em}{\{1, 0.9879324, -0.01, 0.035, 1\}\\\{2, 0.01, $4.837\times10^{-4}$, 0.0009\}}&~~0&- $1.745\times10^{-15}$ - 0.001328$i$&0 - 0.001328$i$\\
			&~~1&$1.553\times10^{-13}$ - 0.001812$i$&0 - 0.001812$i$\\
			&~~2&- $5.712\times10^{-12}$ - 0.002281$i$&0 - 0.002295$i$\\
			\hline
		\end{tabular}
		\caption{\raggedright{The QNMs of near-extremal black string with the negative sign for the square root term. The value of $Q$ is fine tuned such that $r_{Cauchy}\sim r_{H}$ while the value of $k$ and $\alpha_{g}$ is chosen to be 1. $\bar{x}/x_{H}$ is set to 0.8.}}
		\label{QNMPDS-}
	\end{table}
	\begin{table}[h]
		\centering
		\begin{tabular}{|c|m{2em}|c|c|}
			\hline
			\{$M$, $Q$, $\alpha^{2}_{m}$, $\gamma$, $\epsilon$\} \{$\lambda$, $m_{s}$, $\kappa_{H}$, $\bigtriangleup r/r_{H}$\}&~~$n$&AIM~($10^{-5}$)&Formula~(\ref{bseq})~($10^{-5}$)\\
			\hline
			\multirow{3}{14em}{\{1, 0.2, -0.01, -0.08839355, 1\}\\\{0, 0.01, $2.471\times10^{-5}$, 0.00025\}}&~~0&$\pm2.727$ - 1.235$i$&$\pm2.727$ - 1.235$i$\\
			&~~1&$\pm2.727$ - 3.706$i$&$\pm2.727$ - 3.706$i$\\
			&~~2&$\pm2.767$ - 6.195$i$&$\pm2.727$ - 6.177$i$\\
			\hline
			\multirow{3}{14em}{\{1, 0.2, -0.01, -0.08839355, 1\}\\\{1, 0.01, $2.471\times10^{-5}$, 0.00025\}}&~~0&$\pm4.048$ - 1.235$i$&$\pm4.049$ - 1.235$i$\\
			&~~1&$\pm4.048$ - 3.705$i$&$\pm4.049$ - 3.706$i$\\
			&~~2&$\pm4.054$ - 6.156$i$&$\pm4.049$ - 6.177$i$\\
			\hline
			\multirow{3}{14em}{\{1, 0.2, -0.01, -0.08839355, 1\}\\\{2, 0.01, $2.471\times10^{-5}$, 0.00025\}}&~~0&$\pm6.576$ - 1.235$i$&$\pm6.576$ - 1.235$i$\\
			&~~1&$\pm6.576$ - 3.706$i$&$\pm6.576$ - 3.706$i$\\
			&~~2&$\pm6.578$ - 6.180$i$&$\pm6.576$ - 6.177$i$\\
			\hline
			\multirow{3}{14em}{\{1, 0.1, -0.01, -0.08771526, 1\} \{2, 0.01, $3.581\times10^{-5}$, 0.00037\}}&~~0&$\pm9.490$ - 1.790$i$&$\pm9.490$ - 1.790$i$\\
			&~~1&$\pm9.490$ - 5.370$i$&$\pm9.490$ - 5.371$i$\\
			&~~2&$\pm9.483$ - 8.935$i$&$\pm9.490$ - 8.952$i$\\
			\hline
			\multirow{3}{14em}{\{1, 0.5, -0.04, -0.00146486, 1\} \{2, 0.01, $5.572\times10^{-5}$, 0.00034\}}&~~0&$\pm12.591$ - 2.786$i$&$\pm12.592$ - 2.786$i$\\
			&~~1&$\pm12.591$ - 8.357$i$&$\pm12.592$ - 8.358$i$\\
			&~~2&~$\pm12.595$ - 13.925$i$~~&~$\pm12.592$ - 13.929$i$~~\\
			\hline
		\end{tabular}
		\caption{\raggedright{The QNMs of near-extremal black string in units of $10^{-5}$. In this case, the value of $\gamma$ is fine tuned so that $r_{H}\sim r_{\Lambda}$. The physical Universe is extremely small for this limit. Hence, $\bar{x}$ is set to $2x_{H}x_{\Lambda}/(x_{H}+x_{\Lambda})$.}}
		\label{QNMIMS}
	\end{table}

\subsection{QNMs of Near-Extremal Black String in asymptotically AdS}

AIM is used for $r_{C}\sim r_{H}$~(with $r_{\Lambda-}$ and $r_{\Lambda+}$ exist) and $r_{H}\sim r_{\Lambda-}$ cases. While the Spectral Method is used for $r_{C}\sim r_{H}$~(with the absence of $r_{\Lambda-}$ and $r_{\Lambda+}$) and $r_{\Lambda-}\sim r_{\Lambda+}$ cases. The results are shown in TABLE~\ref{QNMPDS+}, \ref{QNMPDS-} and \ref{QNMIMS}.

\section{Spin-1 Perturbations}\label{Sone}

In this section, we demonstrate that spin-1 perturbations in the generic spherically symmetric near-extremal background obey the same analytic formula (\ref{QNMF}). First, there are three orthonormal bases for the spin-1 field.

There are three orthonormal bases,
	\begin{eqnarray}
		\vec{Y}_{\ell m}&=&Y_{\ell m}\hat{r},\nonumber \\
		\vec{\Psi}_{\ell m}&=&\left(\hat{\theta}\frac{\partial Y_{\ell m}}{\partial\theta}+\hat{\varphi}\frac{\partial Y_{\ell m}}{\partial\varphi}\right),\nonumber \\
		\vec{\Phi}_{\ell m}&=&\left(\frac{\hat{\theta}}{\sin\theta}\frac{\partial}{\partial\varphi}Y_{\ell m}-\hat{\varphi}\sin\theta\frac{\partial}{\partial\theta}Y_{\ell m}\right),\nonumber
	\end{eqnarray}
where $Y_{\ell m}$ is the scalar spherical harmonics function \cite{Regge:1957td,Ruffini:1973pta,Hill,Barrera}. The above bases are vector spherical harmonics. Under these bases, any perturbation function can be written as,
	\begin{equation}
		A_{\mu}(t,r,\theta,\varphi)=A^{t}_{\ell m}(t,r)Y_{\ell m}\hat{t}+A^{r}_{\ell m}(t,r)\vec{Y}_{\ell m}+A^{(1)}_{\ell m}(t,r)\vec{\Psi}_{\ell m}+A^{(2)}_{\ell m}(t,r)\vec{\Phi}_{\ell m}.\nonumber
	\end{equation}
The superscript $t,r$ of the gauge field simply defines the components and have nothing to do with spacetime transformation property. Under parity transformation, $\theta\rightarrow\pi-\theta$ and $\varphi\rightarrow\varphi+\pi$, the above expression can be separated into two different parity modes. The odd parity whose parity transformation yields a factor of $(-1)^{\ell+1}$. And the even parity whose parity transformation yields a factor of $(-1)^{\ell}$. Hence, the general representation of vector fields are
	\begin{eqnarray}
		A^{\rm even}_{\mu}(t,r,\theta,\varphi)&=&A^{t}_{\ell m}(t,r)Y_{\ell m}\hat{t}+A^{r}_{\ell m}(t,r)\vec{Y}_{\ell m}+A^{(1)}_{\ell m}(t,r)\vec{\Psi}_{\ell m},\nonumber \\
		A^{\rm odd}_{\mu}(t,r,\theta,\varphi)&=&A^{(2)}_{\ell m}(t,r)\vec{\Phi}_{\ell m}.\nonumber
	\end{eqnarray}

\subsection{Equation of Motion for Odd Perturbation}

The odd perturbation can be expressed in the spherical coordinates as the following
	\begin{equation}
		A^{\rm odd}_{\mu}(t,r,\theta,\varphi)=e^{-i\omega t}A^{(2)}_{\ell m}(r)\left(\frac{\hat{\theta}}{\sin\theta}\frac{\partial}{\partial\varphi}Y_{\ell m}-\hat{\varphi}\sin\theta\frac{\partial}{\partial\theta}Y_{\ell m}\right).\nonumber
	\end{equation}
The electromagnetic field tensor can be calculated by,
	\begin{equation}
		F^{\mu\nu}=g^{\mu\alpha}g^{\nu\beta}\left(\partial_{\alpha}A^{\rm odd}_{\beta}-\partial_{\beta}A^{\rm odd}_{\alpha}\right).\nonumber
	\end{equation}
Hence, the electromagnetic field tensor has the following components up to a factor of $e^{-i\omega t}$,
	\begin{align}
		F^{t\theta}_{\ell m}&=\frac{i\omega}{r^{2}f(r)\sin\theta}A^{(2)}_{\ell m}(r)\partial_{\varphi}Y_{\ell m},&
		F^{t\varphi}_{\ell m}&=-\frac{i\omega}{r^{2}f(r)\sin\theta}A^{(2)}_{\ell m}(r)\partial_{\theta}Y_{\ell m},\nonumber \\
		F^{r\theta}_{\ell m}&=\frac{f(r)}{r^{2}\sin\theta}\frac{dA^{(2)}_{\ell m}}{dr}\partial_{\varphi}Y_{\ell m},&
		F^{r\theta}_{\ell m}&=-\frac{f(r)}{r^{2}\sin\theta}\frac{dA^{(2)}_{\ell m}}{dr}\partial_{\theta}Y_{\ell m},\nonumber \\
		F^{\theta\varphi}_{\ell m}&=\frac{\ell(\ell+1)}{r^{4}\sin\theta}A^{(2)}_{\ell m}Y_{\ell m}.&&\nonumber
	\end{align}

The radial component of the Maxwell's equation ($\nabla_{\nu}F^{r\nu}=0$) is trivially satisfied. The time component is also zero. The $\theta$ and $\varphi$ component are similar to one another with some difference in overall factor. They satisfy the following relation,
	\begin{equation}
		\sin^{2}\theta\frac{\partial Y_{\ell m}}{\partial\theta}\nabla_{\nu}F^{\theta\nu}=\frac{\partial Y_{\ell m}}{\partial \varphi}\nabla_{\nu}F^{\varphi\nu}.\nonumber
	\end{equation}
As such, only a single equation is obtained for the odd parity perturbation. The Maxwell's equation yields,
	\begin{equation}
		\frac{1}{\sin\theta}\frac{\partial Y_{\ell m}}{\partial\varphi}\left\lbrace\frac{d}{dr}\left(f(r)\frac{d}{dr}A^{(2)}_{\ell m}(r)\right)+\frac{\omega^{2}}{f(r)}A^{(2)}_{\ell m}(r)-\frac{\ell (\ell +1)}{r^{2}}A^{(2)}_{\ell m}(r)\right\rbrace=0.
	\end{equation}
Simplified the above Maxwell's equation yields the following radial equation for odd parity perturbation,
	\begin{equation}
		\frac{d^{2}}{dr_{*}^{2}}A^{(2)}_{\ell m}(r)+\left(\omega^{2}-f(r)\frac{\ell (\ell +1)}{r^{2}}\right)A^{(2)}_{\ell m}(r)=0.  \label{odeq}
	\end{equation}

\subsection{Equation of Motion for Even Perturbation}

Similar to the previous case, the even perturbation is written in the spherical coordinates. The result is as shown below,
	\begin{equation}
		A^{\rm even}_{\mu}(t,r,\theta,\varphi)=e^{-i\omega t}A^{t}_{\ell m}(r)Y_{\ell m}\hat{t}+e^{-i\omega t}A^{r}_{\ell m}(r)Y_{\ell m}\hat{r}+e^{-i\omega t}A^{(1)}_{\ell m}(r)\left(\hat{\theta}\frac{\partial Y_{\ell m}}{\partial\theta}+\hat{\varphi}\frac{\partial Y_{\ell m}}{\partial\varphi}\right),\nonumber
	\end{equation}
This expression yields the following electromagnetic field tensor up to a factor of $e^{-i\omega t}$,
	\begin{align}
		F^{tr}_{\ell m}&=\left(\frac{d}{dr}A^{t}_{\ell m}(r)+i\omega A^{r}_{\ell m}(r)\right)Y_{\ell m},&
		F^{t\theta}_{\ell m}&=\frac{1}{r^{2}f(r)}\Big(i\omega A^{(1)}_{\ell m}(r)+A^{t}_{\ell m}(r)\Big)\partial_{\theta}Y_{\ell m},\nonumber \\
		F^{t\varphi}_{\ell m}&=\frac{1}{r^{2}f(r)\sin^{2}\theta}\Big(i\omega A^{(1)}_{\ell m}(r)+A^{t}_{\ell m}(r)\Big)\partial_{\varphi}Y_{\ell m},&
		F^{r\theta}_{\ell m}&=\frac{f(r)}{r^{2}}\left(\frac{d}{dr}A^{(1)}_{\ell m}(r)-A^{r}_{\ell m}(r)\right)\partial_{\theta}Y_{\ell m},\nonumber \\
		F^{r\varphi}_{\ell m}&=\frac{f(r)}{r^{2}\sin^{2}\theta}\left(\frac{d}{dr}A^{(1)}_{\ell m}(r)-A^{r}_{\ell m}(r)\right)\partial_{\varphi}Y_{\ell m}.&&\nonumber
	\end{align}
The $\varphi$ component of the Maxwell's equation is redundant. By defining an auxiliary function $B_{\ell m}(r)$
	\begin{equation}
		i\omega A^{r}_{\ell m}(r)+\frac{d}{dr}A^{t}_{\ell m}(r)\equiv -\frac{\ell (\ell +1)}{r^{2}}B_{\ell m}(r),
	\end{equation}
we can simplify the Maxwell's equations. The time and radial component of the Maxwell's equations become,
	\begin{eqnarray}
		f(r)\frac{d}{dr}B_{\ell m}(r)+\Big(i\omega A^{(1)}_{\ell m}(r)+A^{t}_{\ell m}(r)\Big)&=&0,\nonumber \\
		-\frac{i\omega}{f(r)}B_{\ell m}(r)+\left(A^{r}_{\ell m}(r)-\frac{d}{dr}A^{(1)}_{\ell m}(r)\right)&=&0.\nonumber
	\end{eqnarray}
By eliminating $A^{(1)}_{\ell m}(r)$ from the above equations, the following equation is obtained
	\begin{eqnarray}
		\frac{d}{dr}\left(f(r)\frac{d}{dr}B_{\ell m}(r)\right)+\frac{\omega^{2}}{f(r)}B_{\ell m}(r)-\frac{\ell (\ell +1)}{r^{2}}B_{\ell m}(r)&=&0,\nonumber \\
		\frac{d^{2}}{dr_{*}^{2}}B_{\ell m}(r)+\left(\omega^{2}-f(r)\frac{\ell (\ell +1)}{r^{2}}\right)B_{\ell m}(r)&=&0.    \label{eveq}
	\end{eqnarray}

The above equations (\ref{odeq}) and (\ref{eveq}) would yield the same result as equation (\ref{Initial}) in the near-extremal limit~($\kappa_{\chi}\to 0$) since the extra term containing $f'(r)\sim \kappa_{\chi}\sinh(\kappa_{\chi}r_{*})$ in the scalar case is negligible~(see Ref.~\cite{Burikham:2020dfi} for detailed calculation). Numerical results show excellent agreement with the analytic formula as depicted in Table~\ref{S1dS}, \ref{S1AdS}.

	\begin{table}[h]
		\centering
		\begin{tabular}{|l|m{2em}|c|c|}
			\hline
			~\{$M$, $Q$, $\Lambda$, $\gamma$, $\tau$\} \{$\ell$, $\kappa_{H}$, $\bigtriangleup r/r_{H}$,~Type\}~~&~~$n$&AIM~($10^{-5}$)&Formula~(\ref{QNMF})~($10^{-5}$)\\
			\hline
			\multirow{2}{20em}{~\{1, 0.96109324, 0.03, 0.1, 1\}\\~\{1, $5.925\times10^{-5}$, 0.000087, $r_{H}\sim r_{C}$;~$\omega^{(+)}$\}}&~~0&0 + $5.18094i$&0 + $5.18109i$\\
			&~~1&0 - $0.743313i$&0 - $0.743659i$\\
			\hline
			\multirow{2}{20em}{~\{1, 0.96109324, 0.03, 0.1, 1\}\\~\{1, $5.925\times10^{-5}$, 0.000087, $r_{H}\sim r_{C}$;~$\omega^{(-)}$\}}&~~0&0 - $11.1062i$&0 - $11.1058i$\\
			&~~1&0 - $17.0314i$&0 - $17.0306i$\\
			\hline
			\multirow{2}{20em}{~\{1, 0.96109324, 0.03, 0.1, 1\}\\~\{2, $5.925\times10^{-5}$, 0.000087, $r_{H}\sim r_{C}$;~$\omega^{(+)}$\}}&~~0&0 + $10.5058i$&0 + $10.5060i$\\
			&~~1&0 + $4.58068i$&0 + $4.58124i$\\
			\hline
			\multirow{2}{20em}{~\{1, 0.96109324, 0.03, 0.1, 1\}\\~\{2, $5.925\times10^{-5}$, 0.000087, $r_{H}\sim r_{C}$;~$\omega^{(-)}$\}}&~~0&0 - $16.4311i$&0 - $16.4307i$\\
			&~~1&0 - $22.3563i$&0 - $22.3555i$\\
			\hline
			\multirow{2}{20em}{~\{1, 0.1, 0.03, -0.08771526, 1\}\\~\{1, $3.581\times10^{-5}$, 0.00037, $r_{H}\sim r_{\Lambda}$;~$\omega^{(\pm)}$\}}&~~0&~$\pm5.839$ - $1.790i$~~&~$\pm5.839$ - $1.790i$~~\\
			&~~1&$\pm5.840$ - $5.371i$&$\pm5.839$ - $5.371i$\\
			\hline
			\multirow{2}{20em}{~\{1, 0.1, 0.03, -0.08771526, 1\}\\~\{2, $3.581\times10^{-5}$, 0.00037, $r_{H}\sim r_{\Lambda}$;~$\omega^{(\pm)}$\}}&~~0&$\pm10.42$ - $1.790i$&$\pm10.43$ - $1.790i$\\
			&~~1&$\pm10.42$ - $5.370i$&$\pm10.43$ - $5.371i$\\
			\hline
		\end{tabular}
		\caption{\raggedright The spin-1 QNMs of near-extremal dS black hole in various types in units of $10^{-5}$. The $\omega^{(+)}$ refers to family of QNMs with a positive sign for the square root term. The $\omega^{(-)}$ is a family of QNMs whose sign of the square root term is negative.}
		\label{S1dS}
	\end{table}
	\begin{table}[h]
		\centering
		\begin{tabular}{|c|m{2em}|c|c|}
			\hline
			~\{$M$, $Q$, $\Lambda$, $\gamma$, $\tau$, $\ell$, $\kappa_{\chi}$\} \{$\bigtriangleup r/r_{\chi}$,~Type\}~~&~~$n$&Numerical~($10^{-6}$)&Formula~(\ref{QNMF})~($10^{-6}$)\\
			\hline
			\multirow{2}{21em}{\{1, 1.0639926, -0.0015, -0.1, 1, 1, $2.052\times10^{-4}$\}\\\{0.00079, $r_{H}\sim r_{C}\&\exists~r_{\Lambda+},r_{\Lambda-}$;~$\omega^{(+)}$\}}&~~0&0 + $275.191i$&0 + $275.252i$\\
			&~~1&0 + $69.9982i$&0 + $70.0910i$\\
			\hline
			\multirow{2}{21em}{\{1, 1.0639926, -0.0015, -0.1, 1, 1, $2.052\times10^{-4}$\}\\\{0.00079, $r_{H}\sim r_{C}\&\exists~r_{\Lambda+},r_{\Lambda-}$;~$\omega^{(-)}$\}}&~~0&0 - $480.499i$&0 - $480.413i$\\
			&~~1&0 - $685.537i$&0 - $685.574i$\\
			\hline
			\multirow{2}{21em}{\{1, 1.0639926, -0.0015, -0.1, 1, 2, $2.052\times10^{-4}$\}\\\{0.00079, $r_{H}\sim r_{C}\&\exists~r_{\Lambda+},r_{\Lambda-}$;~$\omega^{(+)}$\}}&~~0&0 + $535.442i$&0 + $535.539i$\\
			&~~1&0 + $330.301i$&0 + $330.378i$\\
			\hline
			\multirow{2}{21em}{\{1, 1.0639926, -0.0015, -0.1, 1, 2, $2.052\times10^{-4}$\}\\\{0.00079, $r_{H}\sim r_{C}\&\exists~r_{\Lambda+},r_{\Lambda-}$;~$\omega^{(-)}$\}}&~~0&0 - $740.780i$&0 - $740.700i$\\
			&~~1&0 - $946.117i$&0 - $945.861i$\\
			\hline
			\multirow{2}{21em}{\{1, 0.9576851, -0.0015, 0.1, 1, 1, $4.038\times10^{-4}$\}\\\{0.00056, $r_{H}\sim r_{C}\&\nexists~r_{\Lambda+},r_{\Lambda-}$;~$w^{(-)}$\}}&~~0&0 - $660.603i$&0 - $660.466i$\\
			&~~1&0 - $1017.26i$&0 - $1016.93i$\\
			\hline
			\multirow{2}{21em}{\{1, 0.9576851, -0.0015, 0.1, 1, 2, $2.052\times10^{-4}$\}\\\{0.00056, $r_{H}\sim r_{C}\&\nexists~r_{\Lambda+},r_{\Lambda-}$;~$w^{(-)}$\}}&~~0&0 - $974.671i$&0 - $974.545i$\\
			&~~1&0 - $1331.34i$&0 - $1331.01i$\\
			\hline
			\multirow{2}{21em}{\{1, 0.3, -0.015, -0.147034169, 1, 1, $6.186\times10^{-6}$\}\\\{0.00015, $r_{H}\sim r_{\Lambda-}$;~$\omega^{(\pm)}$\}}&~~0&$\pm14.44$ - $3.092i$&$\pm14.44$ - $3.093i$\\
			&~~1&$\pm14.44$ - $9.211i$&$\pm14.44$ - $9.279i$\\
			\hline
			\multirow{2}{21em}{\{1, 0.3, -0.015, -0.147034169, 1, 2, $6.186\times10^{-6}$\}\\\{0.00015, $r_{H}\sim r_{\Lambda-}$;~$\omega^{(\pm)}$\}}&~~0&$\pm25.39$ - $3.093i$&$\pm25.39$ - $3.093i$\\
			&~~1&$\pm25.39$ - $9.275i$&$\pm25.39$ - $9.279i$\\
			\hline
			\multirow{2}{22em}{\{1, 1, -0.0015, -0.0436850385, 1, 1, $6.652\times10^{-7}$\}\\\{0.000066, $r_{\Lambda+}\sim r_{\Lambda-}$;~$w^{(-)}$\}}&~~0&0 - $59.6096i$&0 - $59.6051i$\\
			&~~1&0 - $87.9309i$&0 - $87.9518i$\\
			\hline
			\multirow{2}{22em}{\{1, 1, -0.0015, -0.0436850385, 1, 2, $6.652\times10^{-7}$\}\\\{0.000066, $r_{\Lambda+}\sim r_{\Lambda-}$;~$w^{(-)}$\}}&~~0&0 - $90.2776i$&0 - $90.2678i$\\
			&~~1&0 - $118.569i$&0 - $118.614i$\\
			\hline
		\end{tabular}
		\caption{\raggedright The spin-1 QNMs of near-extremal black hole in various asymptotically AdS background in units of $10^{-6}$. The $\omega^{(+)}$ refers to family of QNMs with a positive sign for the square root term. The $\omega^{(-)}$ is a family of QNMs whose sign of the square root term is negative. $w^{(-)}$ indicates the QNMs obtained from the Spectral Method.}
		\label{S1AdS}
	\end{table}

\section{Discussions and Conclusions}\label{conclude}

The QNMs of neutral spin-0 scalar field in the near-extremal black holes/strings in generic spherically/cylindrically symmetric background are calculated using approximate analytic formula and compared with two numerical methods, AIM and the spectral method. Both analytic and numerical results are in good agreement. The analytic formula becomes more accurate as the black hole/string approaches extremality. The analytic formula is used to explore the Strong Cosmic Censorship conjecture in Ref.~\cite{Burikham:2020dfi}. It is also shown that the QNMs of spin-1 field in the similar near-extremal background obey the same analytic formula for both even and odd parity.

\acknowledgments

T.W. is financially supported by Chulalongkorn University, Dutsudi Phiphat Scholarship. S.~P. is supported by Rachadapisek Sompote Fund for Postdoctoral Fellowship, Chulalongkorn University. P.B. is supported in part by the Thailand Research Fund (TRF), Office of Higher Education Commission (OHEC) and Chulalongkorn University under grant RSA6180002. 

\appendix

\section{Scaling of Parameters}\label{ReS}

Throughout this paper the value of $M$ is set to unity. However, it is possible to rescale $M$ to any other values. This can be achieved by rescaling $r=M\tilde{r}$ and rewriting the metric function, spacetime parameters and the Klein-Gordon equation in terms of $\tilde{r}$ and rescaled quantities. These transformations are listed in Table \ref{RP}.
	\begin{table}[h]
		\centering\
		\begin{tabular}{|c|c|c|}
			\hline
			Quantity&~General Spacetime Parameter~~&~The Tilde Notation~~\\
			\hline
			Mass&$M$&$1$\\
			\hline
			Charge&$Q$&$\tilde{Q}=Q/M$\\
			\hline
			~Graviton Self-interaction parameter~~&$\gamma$&$\tilde{\gamma}=\gamma M$\\
			\hline
			Cosmology Constant&$\Lambda$&$\tilde{\Lambda}=\Lambda M^{2}$\\
			\hline
			Scalar Mass&$m_{s}$&$\tilde{m}_{s}=m_{s}M$\\
			\hline
			Quasinormal Frequency&$\omega$&$\tilde{\omega}=\omega M$\\
			\hline
		\end{tabular}
		\caption{Table of rescaled parameters}
		\label{RP}
	\end{table}

\section{Asymptotic Iteration Method}\label{SectAIM}

Throughout the asymptotically dS and some of the AdS cases in this work, AIM is used to numerically calculate QNMs of black holes and black strings. In this section, AIM is elaborated in details. 

For a generic radial wave equation,
	\begin{equation}
		f\phi''+f'\phi'+\left(\frac{\omega^{2}}{f}-\frac{\ell(\ell +1)}{r^{2}}-\frac{f'}{r}-m^{2}_{s}\right)\phi=0,
	\end{equation}
where $f'=df/dr$. For convenience, a new variable $x=1/r$ is used. The above equation becomes,
	\begin{equation}
		\phi''+\frac{p'}{p}\phi'+\left[\left\lbrace\frac{\omega}{p}\right\rbrace^{2}-\frac{1}{p}\left(\ell(\ell +1)+2Mx-2Q^{2}x^{2}+\frac{\gamma}{x}-\frac{2\Lambda}{3x^{2}}+\frac{m^{2}_{s}}{x^{2}}\right)\right]\phi=0,
	\end{equation}
where
	\begin{equation}
		p=Q^{2}x^{4}-2Mx^{3}+\tau x^{2}+\gamma x-\frac{\Lambda}{3}.
	\end{equation}
Here and henceforth $'$ represents $d/dx$ . The metric function $f(x)$ can be written as,
	\begin{equation}
		f(x)=\frac{1}{x^{2}}(x-x_{1})(x-x_{2})(x-x_{3})(x-x_{4}).\nonumber
	\end{equation}
The surface gravity now becomes
	\begin{equation}
		\kappa_{i}=\left.\frac{1}{2}\frac{df}{dr}\right|_{r=r_{i}}=-\frac{x^{2}}{2}f'(x_{i})=-\frac{1}{2}\prod_{j\neq i}(x_{i}-x_{j}).\nonumber
	\end{equation}
We then introduce the tortoise coordinate $dr_{*}=dr/f$. Thus we can expressed tortoise coordinate in term of $x$
\begin{equation}
		r_{*}=-\int\frac{dx}{x^{2}f(x)}
		=-\int\frac{dx}{(x-x_{1})(x-x_{2})(x-x_{3})(x-x_{4})}
	=-\int dx\sum^{4}_{i=1}\frac{A_{i}}{(x-x_{i})}.\nonumber
	\end{equation}
The last step in the equation above implies 
	\begin{eqnarray}
		1&=&\sum^{4}_{i=1}\left(A_{i}\prod_{j\neq i}(x-x_{j})\right).\nonumber
	\end{eqnarray}
By substituting $x$ with each root of $f(x)$, it is clear that $A_{i}=-1/2\kappa_{i}$. Thus,
	\begin{eqnarray}
		r_{*}&=&\int dx\sum^{4}_{i=1}\frac{1}{2\kappa_{i}(x-x_{i})},\nonumber \\
		&=&\ln\left(\prod^{4}_{i=1}(x-x_{i})^{1/2\kappa_{i}}\right).\nonumber \\
		\therefore e^{i\omega r_{*}}&=&\prod^{4}_{i=1}(x-x_{i})^{i\omega/2\kappa_{i}}.
	\end{eqnarray}
In order to scale out the divergent behaviour at the cosmic horizon, the wave function is defined as follows
	\begin{equation}
		\phi(x)=e^{i\omega r_{*}}u(x).\nonumber
	\end{equation}
The wave equation then takes the following form,
	\begin{equation}
		u''+\frac{p'-2i\omega}{p}u'-\frac{1}{p}\left[\ell(\ell +1)+2Mx-2Q^{2}x^{2}+\frac{\gamma}{x}-\frac{2\Lambda}{3x^{2}}+\left(\frac{m_{s}}{x}\right)^{2}\right]u=0.
	\end{equation}
Once more the divergent behavior is needed to be scaled out. At the event horizon, the scaling is done by defining the following function,
	\begin{equation}
		u(x)=(x-x_{1})^{-i\omega/2\kappa_{1}}\chi(x),\nonumber
	\end{equation}
where the subscript `$1$' indicates that the parameter is associated with the event horizon. 

Ultimately, the radial wave equation can be cast in the following form,
	\begin{equation}
		\chi''(x)=\lambda_{0}(x)\chi'(x)+s_{0}(x)\chi(x), \label{B6}
	\end{equation}
where
	\begin{eqnarray}
		\lambda_{0}&=&-\frac{4i\omega}{Q^{2}\prod_{i=1}^{4}(x-x_{i})}-\frac{p'-2i\omega}{p},\nonumber \\
		s_{0}&=&\frac{\ell(\ell +1)+2x(M-Q^{2}x)}{p}+\frac{2m^{2}_{s}+3\gamma x+2\Lambda}{3px^{2}}\nonumber \\
		&&-\frac{2i \omega p'+4\omega^{2}}{pQ^{2}(x-x_{1})\prod^{4}_{i=2}(x_{1}-x_{i})}+\left(\frac{2\omega}{Q^{2}(x-x_{1})\prod^{4}_{i=2}(x_{1}-x_{i})}\right)^{2}\nonumber \\
		&&+\frac{2i\omega}{Q^{2}(x-x_{1})^{2}\prod^{4}_{i=2}(x_{1}-x_{i})}.
	\end{eqnarray}
By differentiating (\ref{B6}) with respect to $x$ for $n$ times. We obtain
	\begin{equation}
		\chi^{(n)}=\lambda_{n-2}\chi'+s_{n-2}\chi,
	\end{equation}
where
	\begin{eqnarray}
		\lambda_{n}&=&\lambda_{n-1}'+\lambda_{n-1}\lambda_{0}+s_{n-1},\nonumber \\
		s_{n}&=&s_{n-1}'+s_{0}\lambda_{n-1}.\nonumber
	\end{eqnarray}
The asymptotic behavior of $\chi^{(n)}$ implies that,
	\begin{equation}
		\frac{s_{n}}{\lambda_{n}}\equiv\beta,\nonumber
	\end{equation}
where $\beta$ is some constant. Hence, the following relation is obtained,
	\begin{equation}
		\lambda_{n}s_{n-1}=\lambda_{n-1}s_{n}.
	\end{equation}
In order to calculate the energy eigenvalues, the radial equation must be solved. Assuming that $n$ is sufficiently large such that the above relation holds, the following equation is obtained \cite{Ciftci:2003mp},
	\begin{eqnarray}
		\frac{\chi^{(n+2)}}{\chi^{(n+1)}}&=&\frac{\lambda_{n}(\chi'+\beta\chi)}{\lambda_{n-1}(\chi'+\beta\chi)},\nonumber \\
		\frac{d}{dx}\ln\chi^{(n+1)}&=&\frac{\lambda_{n-1}'+\lambda_{n-1}\lambda_{0}+s_{n-1}}{\lambda_{n-1}}=(\ln\lambda_{n-1})'+\lambda_{0}+\beta.\nonumber \\
		\therefore\chi^{(n+1)}&=&A\lambda_{n-1}e^{\int(\lambda_{0}+\beta)dx},
	\end{eqnarray}
where $A$ is an integration constant. From the recursive relation, $\chi^{(n+1)}/\lambda_{n-1}=\chi'+\beta\chi$. Hence
	\begin{eqnarray}
		\chi'+\beta\chi&=&Ae^{\int(\lambda_{0}+\beta)dx},\nonumber \\
		\frac{1}{e^{\beta x}}(e^{\beta x}\chi)'&=&Ae^{\int\lambda_{0}dx+\beta x},\nonumber \\
		\chi&=&e^{-\beta x}\left\lbrace B+A\int e^{\int\lambda_{0}dx+2\beta x}dx\right\rbrace.
	\end{eqnarray}
With the wave function~$\chi(x)$ known, it is straightforward to calculate the energy eigenvalues. Each coefficient of $\lambda_{i}$, $s_{i}$ and their derivatives must be computed for all iterations. This feature becomes the main disadvantage of the method. To alleviate this problem, an improved version of the method is proposed in \cite{Cho:2009cj}. By expanding $\lambda_{n}$ and $s_{n}$ around some arbitrary point $\bar{x}$ in the Taylor series, the necessity of computing derivative at each iteration is eliminated. However, it is important to ensure that $\bar{x}$ is set within the physical spacetime under consideration. The coefficient is now expressed as follows \cite{Cho:2009cj},
	\begin{eqnarray}
		\lambda_{n}(x)&=&\sum^{\infty}_{i=0}c^{i}_{n}(x-\bar{x})^{i}, \\
		s_{n}(x)&=&\sum^{\infty}_{i=0}d^{i}_{n}(x-\bar{x})^{i}.
	\end{eqnarray}
The recursive relations for $\lambda_{n}$ and $s_{n}$ become,
	\begin{eqnarray}
		c^{i}_{n}&=&(i+1)c^{i+1}_{n-1}+d^{i}_{n-1}+\sum^{\infty}_{j=0}c^{j}_{0}c^{i-j}_{n-1}, \\
		d^{i}_{n}&=&(i+1)d^{i+1}_{n-1}+\sum^{\infty}_{j=0}d^{j}_{0}c^{i-j}_{n-1}.
	\end{eqnarray}
The asymptotic behavior is now taking the following form,
	\begin{equation}
		d^{0}_{n}c^{0}_{n-1}-d^{0}_{n-1}c^{0}_{n}=0.
	\end{equation}
The quasinormal frequencies can be calculated by computing $c^{i}_{n}$ and $d^{i}_{n}$ starting from $n=0$ until it reaches a desirable number of recursions, where the asymptotic behavior is sufficiently satisfied. The wave function $\chi(x)$ is then calculated. Throughout this paper, all asymptotically dS background are calculated via the improved AIM as well as some cases of the asymptotically AdS background.

\section{The Spectral Method}\label{SM}

This method is used for an unbounded physical Universe whose domain of $r$ is between $[r_{H},+\infty)$, where $r_{H}$ is the event horizon of any given black hole. The boundary condition at the AdS infinity must be the diminishing waves while only ingoing waves $e^{-i\omega r_{*}}$ are allowed at the horizon. The equation of motion is thus linearized by setting $\phi=e^{-i\omega r_{*}}S$. Additionally, we set $u=r_{H}/r$ so that the domain of physical space is $u \in [0,1]$. The equation of motion becomes,
	\begin{equation}
		u^{2}\frac{\partial}{\partial u}\left(f(u)u^{2}\frac{\partial}{\partial u}S(u)\right)+2iw u^{2}\frac{\partial}{\partial u}S(u)-r^{2}_{H}V_{eff}(u)S(u)=0,
	\end{equation}
where $w=\omega r_{H}$ and other parameters are scaled accordingly; $m_{s}\to m_{s}r_{H}, M\to M/r_{H}, Q^{2}\to Q^{2}/r_{H}^{2},\gamma\to \gamma r_{H}, \Lambda\to \Lambda r_{H}^{2}$. Since the solution is regular at both boundaries, the solution can be expressed as,
	\begin{equation}
		S(u)=\sum^{N}_{n=0}b_{n}T_{n}(2u-1),
	\end{equation}
where $T_{n}$ is the Chebyshev polynomials of the first kind. The domain $2u-1$ is used in order to change the domain of interest to $[-1,1]$. The expansion will become more accurate with larger $N$ and be exact in $N\to \infty$ limit due to the completeness of orthonormal Chebyshev polynomials. We use the Gauss-Lobatto grid 
\be
u_{k} = \frac{1}{2}\Big( 1+\cos\left( \frac{k\pi}{N}\right)\Big),
\ee
where $k=0,1,2,.., N$ to transform the equations into the generalized eigenvalue problem. The linear equations can then be solved for the coefficient $b_{n}$ and $w$ for any $N$. For sufficiently large $N=300$, the value of $w$ converges with great accuracy~\cite{Burikham:2017gdm,Ponglertsakul:2018smo}.

\end{document}